\documentclass{aa}  
\usepackage{graphicx}
\usepackage{hyperref}
\hypersetup{
    colorlinks=true,
    linkcolor=blue,
    citecolor=blue,
    filecolor=magenta,      
    urlcolor=blue,
}
\usepackage{xcolor}
\usepackage{txfonts}
\usepackage{natbib}
\bibpunct{(}{)}{;}{a}{}{,} 
\usepackage{tikz} 
\usetikzlibrary{shapes, arrows} 
\usepackage{adjustbox} 
\usepackage{centernot} 
\usepackage{subcaption} 
\usepackage{appendix} 
\usepackage{float} 

\begin{document} 

\title{The interior of Uranus}
\subtitle{Thermal profile, bulk composition, and the distribution of rock, water, and hydrogen and helium}

\author{Luca Morf
        \and
        Simon Müller
        \and
        Ravit Helled
        }

\institute{Department of Astrophysics, University of Z\"urich,
          Winterthurerstrasse 190, 8057 Z\"urich, Switzerland\\
          \email{luca.morf@uzh.ch}
         }

\date{Received 13 May 2024; accepted 16 August 2024}

    
\abstract{
    We present improved empirical density profiles of Uranus and interpret them in terms of their temperature and composition using a new random algorithm.
    The algorithm to determine the temperature and composition is agnostic with respect to the temperature gradient in non-isentropic regions and chooses amongst all possible gradients randomly that are stable against convection and correspond to an Equation of State (EoS) compatible composition. 
    Our empirical models are based on an efficient implementation of the Theory of Figures (ToF) up to tenth order including a proper treatment of the atmosphere.
    The accuracy of tenth order ToF enables us to present accurate calculations of the gravitational moments of Uranus up to $J_{14}$: 
    $J_{6}  = ( 5.3078 \pm 0.3312)\cdot10^{-7}$,
    $J_{8}  = (-1.1114 \pm 0.1391)\cdot10^{-8}$,
    $J_{10} = ( 2.8616 \pm 0.5466)\cdot10^{-10}$,
    $J_{12} = (-8.4684 \pm 2.0889)\cdot10^{-12}$, and
    $J_{14} = ( 2.7508 \pm 0.7944)\cdot10^{-13}$. 
    We consider two interior models of Uranus that differ with respect to the maximal number of materials allowed per layer of Uranus (three versus four composition components). The case with three materials does not allow Hydrogen and Helium (H-He) in deeper parts of Uranus and results in a higher water (H$_{2}$O) abundance which leads to lower central temperatures. On the other hand, the models with four materials allow H-He to be mixed into the deeper interior and lead to rock-dominated solutions. We find that these four composition components' models are less reliable due to the underlying empirical models' incompatibility with realistic Brunt frequencies.
    Most of our models are found to be either purely convective with the exception of boundary layers, or only convective in the outermost region above $\sim 80 \%$ of the planets' radius $r_{U}$. Almost all of our models possess a region ranging between $\sim(0.75-0.9)r_{U}$ that is convective and consists of ionic H$_{2}$O which could explain the generation of Uranus' magnetic field.
    
    }

\maketitle

\section{Introduction}
\label{sec:Introduction}

Uranus has only been visited by the Voyager 2 spacecraft. This Uranus flyby yielded valuable gravity data \citep[for example][]{French_1988} that are used to constrain the planetary interior. However, since it was a single flyby, only the first two gravitational moments ($J_{2}$ and $J_{4}$) were measured with rather high uncertainties. These values have been later improved using observations of Uranus'  moons and rings \citep[for example][]{Jacobson_2014,French_2024}.  Nonetheless, Uranus' gravity data remain  insufficient to constrain its internal structure and bulk composition. 
\par 
At the moment, various key open questions regarding Uranus' nature remain open. For example, the name 'ice giant' may not be appropriate for Uranus, since it remains unknown whether Uranus' bulk composition is water- or rock-dominated \citep{Helled_2020_2}. Another unsolved mystery is the origin of Uranus' measured low heat flux. One possibility is that Uranus' deep interior is still very hot and that the primordial heat cannot escape due to thermal boundary layers or non-convective regions caused by composition gradients \cite[for example][]{Nettelmann_2016, Vazan_2020,Scheibe_2019,Scheibe_2021}. Alternatively, it is possible that Uranus' low luminosity corresponds to a cold interior, which could be a result of a giant impact that led to rapid cooling (\cite{Reinhardt_2020}). Improving our understanding of Uranus' interior and composition could aid in testing and constraining planet formation and evolution models. 
\par
Interior models are designed to fit the measured physical properties of the planets. These correspond to the planetary mass, radius, rotation rate, and gravitational field (gravitational moments). There are two widely used methods to model the planetary internal structure. The first corresponds to the standard approach, the so-called physical interior models, where the structure equations are solved together with an Equation of State (EoS) of the materials assumed by the modeller. This includes assuming a heat transport mechanism within the planet to get a temperature profile of the planetary interior. In the simplest case, the temperature profile is assumed to be adiabatic corresponding to a fully convective interior. Further details on interior modelling can also be found in \cite{Helled_2020_1} and the references therein. 
\par
The second method, the so-called empirical interior models, infers density profiles compatible with the measurements without making an a piori assumption on the planetary composition and temperature profile. Density profiles are typically represented by polytropes \citep[for example][]{Neuenschwander_2021,Neuenschwander_2022}, polynomials \citep[for example][]{Helled_2010, Movshovitz_2022}, or by an arbitrary function \citep[for example][]{Marley_1995,Podolak_2000,Podolak_2022}. The pressure is inferred in a second step assuming hydrostatic equilibrium. Since no assumptions are made for the materials and heat transport mechanism, the composition and the temperature profile of the planet remain unknown. This approach finds significantly more density profiles than physical models, but may also include nonphysical ones. For example, empirical density profiles can posses a central density that is too high for any plausible material to sustain.
\par
In general, determining the composition and internal structure of a planet is challenging due to the degenerate  nature of this problem. Even for Jupiter, whose planetary mass, radius, rotation rate, gravitational field, and atmospheric composition have been accurately measured, the data are insufficient to uniquely determine Jupiter's interior \citep[for example][]{Miguel_2022,Howard_2023,Militzer_2024}, regardless of the method used. 
\par
For Uranus, the available data are limited in comparison to the gas giant planets. However, several studies have investigated Uranus' internal structure in detail.  For example, \cite{Nettelmann_2013} presented physical interior models of Uranus (and Neptune) where the interior was divided into three layers: a rocky centre surrounded by two adiabatic and homogeneous envelopes of hydrogen, helium, and water. They found the outer envelope metallicity to be $Z\leq8\%$  for modified shape and rotation data and $Z\leq17\%$ for original values provided by Voyager 2. In each of these cases, solar metallicity models ($Z=0.015$) were still found to be possible. \cite{Neuenschwander_2022} presented  empirical interior models of Uranus and Neptune. It was shown that faster rotation and/or deep winds favour centrally concentrated density distributions. It was also argued that a more accurate determination of the gravity field can significantly reduce the possible range of internal structures. In addition, \cite{Neuenschwander_2022}  predicted the values of $J_6$ and $J_8$ for both planets based on their model. Since information about composition and temperature is crucial to get a complete understanding of the planet, several algorithms have been proposed to interpret empirical density profiles in terms of temperature and composition \citep[for example][]{Podolak_2022, Neuenschwander_2024}.
\par
\cite{Podolak_2022} presented a method that combines a top-down with a down-up approach. For the top-down part, the algorithm starts at the planetary surface using the expected temperature and composition values of Uranus at the 1-bar pressure level. It then divides the planets into shells and moves inwards. At each consecutive shell, the algorithm attempts to minimise the mass fraction of high-Z material whenever possible (first water, then rock, then iron) while keeping the temperature constant. If no mixture matches the corresponding density and pressure, the temperature is increased while keeping the same composition. This procedure results in a composition that favours the material with the lowest mean molecular weight and keeps the temperature as low as possible. For the bottom-up part, the algorithm starts with a postulated temperature and composition of the centre. Continuing outwards, the algorithm checks whether the same composition matches the density and pressure while keeping the temperature monotonically decreasing. If the check fails, a composition is inferred that retains the previous temperature by decreasing the metal content (first iron, then rock, then water). The top-down and bottom-up algorithms were combined by comparing the inferred composition profiles, and choosing the solution which minimises the amount of water (layer by layer). Using this method,  \cite{Podolak_2022} concluded that the water-to-rock ratio in Uranus cannot be much smaller than 0.5. \par
The algorithm presented in \cite{Neuenschwander_2024} is based on a top-down approach. It begins  with the planetary surface values that include Uranus' atmospheric composition and temperature. Then, the algorithm attempts to move inwards, assuming a constant composition and entropy. If this fails in some layer, the algorithm uses the Ledoux criterion \citep[][also see section \ref{sec:Method_Algorithm}]{Ledoux_1947} by assuming the fluid to be precisely at the Ledoux limit: the temperature gradient is compensated exactly by the composition gradient to yield neutral buoyancy. The algorithm therefore finds a unique temperature-composition profile for a given density profile. \cite{Neuenschwander_2024} concluded that their empirical density profiles can be interpreted as having non-adiabatic interiors with composition gradients. Compared to fully adiabatic models, these solutions have higher central temperatures. The inferred water-to-rock ratio was found to be between 2.6 and 21.
\par
In this paper, we present a new random algorithm to determine the composition-temperature profiles of improved empirical structure models. Introducing randomness yields a broader range of solutions which is more in line with the spirit of using empirical models. We improve our empirical structure models by calculating the Theory of Figures (\cite{Zharkov_1978}) up to tenth order. Our Theory of Figures (ToF) implementation includes the possibility of incorporating barotropic differential rotation. We also employ better approximations than previous ToF implementations which lead  to a significantly enhanced accuracy. Finally, we include a treatment of the atmosphere which leads to more realistic outermost parts of Uranus in our empirical structure models.
\par
Our paper is structured as follows: In Section \ref{sec:Method} we explain the newly developed methods, such as the inclusion of the atmosphere, the development of a tenth-order ToF, and the new random algorithm that is used to infer the composition, temperature, and stability against convection throughout the planet. We present our results in Section \ref{sec:Results}. 
A comparison of our results with previous studies and a discussion on the limitations of this work are presented in Section \ref{sec:Discussion}. A summary and conclusions are presented in Section \ref{sec:SummaryAndConclusion}.


\section{Methods}
\label{sec:Method}

Our empirical models are based on polytropes which relate the density $\rho$ with the pressure $P$ inside the planet via:
\begin{equation}
    \label{eq:method1}
    \rho(P) = \left(\frac{P}{K}\right)^{\frac{n}{n+1}},
\end{equation}
where $K$ and $n$ are constants. Following \cite{Neuenschwander_2022}, we begin with three consecutive polytropes defined by constants $K_{1, 2, 3}$ and $n_{1, 2, 3}$, where the polytropes are separated by two transition radii. Additionally, we added a fourth polytrope representing the outermost region of the planet which is consistent with atmospheric models of Uranus. Specifically, in the outermost region (atmosphere) we impose the following polytropic relation: 
\begin{equation}
    \label{eq:Atmos1}
    \rho(r \geq r_{*}) = \left(\frac{P(r \geq r_{*})}{K_{\mathrm{atmos}}}\right)^{\frac{n_{\mathrm{atmos}}}{n_{\mathrm{atmos}}+1}},
\end{equation} 
where $K_{\mathrm{atmos}}$ and $n_{\mathrm{atmos}}$ are fitted to the atmosphere model taken from \cite{Hueso_2020}. $P$ is the atmospheric pressure profile obtained by numerically integrating the condition for hydrostatic equilibrium, $\rho(r)^{-1}\vec{\nabla}P(r) = \vec{\nabla}U(r)$, where $U$ is the total potential (gravity and rotation). We let $r_{*}$ to be freely chosen by the algorithm as long as $P(r_{*}) \geq P_{\mathrm{min}}$ and set $P_{\mathrm{min}} = 100$ bar. In order to ensure mass conservation we applied: 
\begin{equation}
    \label{eq:Atmos2}
    \rho(r) \rightarrow \frac{\rho(r)}{4 \pi \int_{0}^{r_{\mathrm{max}}} r^2 \rho(r) \mathrm{d}r}\cdot M, 
\end{equation}
after using equation \ref{eq:Atmos1} until convergence was accomplished. 
\par
The polytropic relation between the density $\rho$ and the pressure $P$ in equation \ref{eq:method1} is only inspired by, not based on physical materials. The constants $K_{1, 2, 3}$ and $n_{1, 2, 3}$ are chosen to exclusively fit the gravity data regardless of any material properties, and therefore the density profiles are empirical. As we discuss in section \ref{sec:Method_Algorithm}, the inferred $P-\rho$ relations are then interpreted in terms of composition and temperature using  physical Equations of State (EoS).
\par 
We employed the Markov-Chain-Monte-Carlo (MCMC) \texttt{emcee} algorithm (\cite{Foreman-Mackey_2013}) and used the ToF (\cite{Zharkov_1978}) with $N=4096$ spheroids and no interpolation to find empirical profiles of Uranus. They assume uniform rotation and agree within 1$\sigma$ with the measured gravitational moments listed in Table \ref{tab:uranus_data}. 

\begin{table}[H]
\centering            
\caption{Physical parameters used for constraining our 4-polytrope empirical models of Uranus.} 
\begin{tabular}{l|l|l}
\hline
$G$                & $6.6743  \cdot 10^{-11}$  N m$^{2}$ kg$^{-2}$ & \cite{Tiesinga_2021} \\
$M$                & $8.68127 \cdot 10^{25}$   kg                  & \cite{Jacobson_2014} \\
$R_{\mathrm{eq}}$  & $2.5559  \cdot 10^{7}$    m                   & \cite{Lindal_1987} \\
$P_{\mathrm{rot}}$ & $62080$                   s                   & \cite{Desch_1986} \\
$P_{0}$            & $10^{5}$                  Pa                  &  \\
$J_{2}$            & $(3510.68 \pm 0.7) \cdot 10^{-6}$             & \cite{Jacobson_2014} \\
$J_{4}$            & $(-34.17  \pm 1.3) \cdot 10^{-6}$             & \cite{Jacobson_2014}\\
\hline
\end{tabular}%

\label{tab:uranus_data} 
\end{table}

\begin{figure}[H]
    \centering
    \includegraphics[width=\hsize]{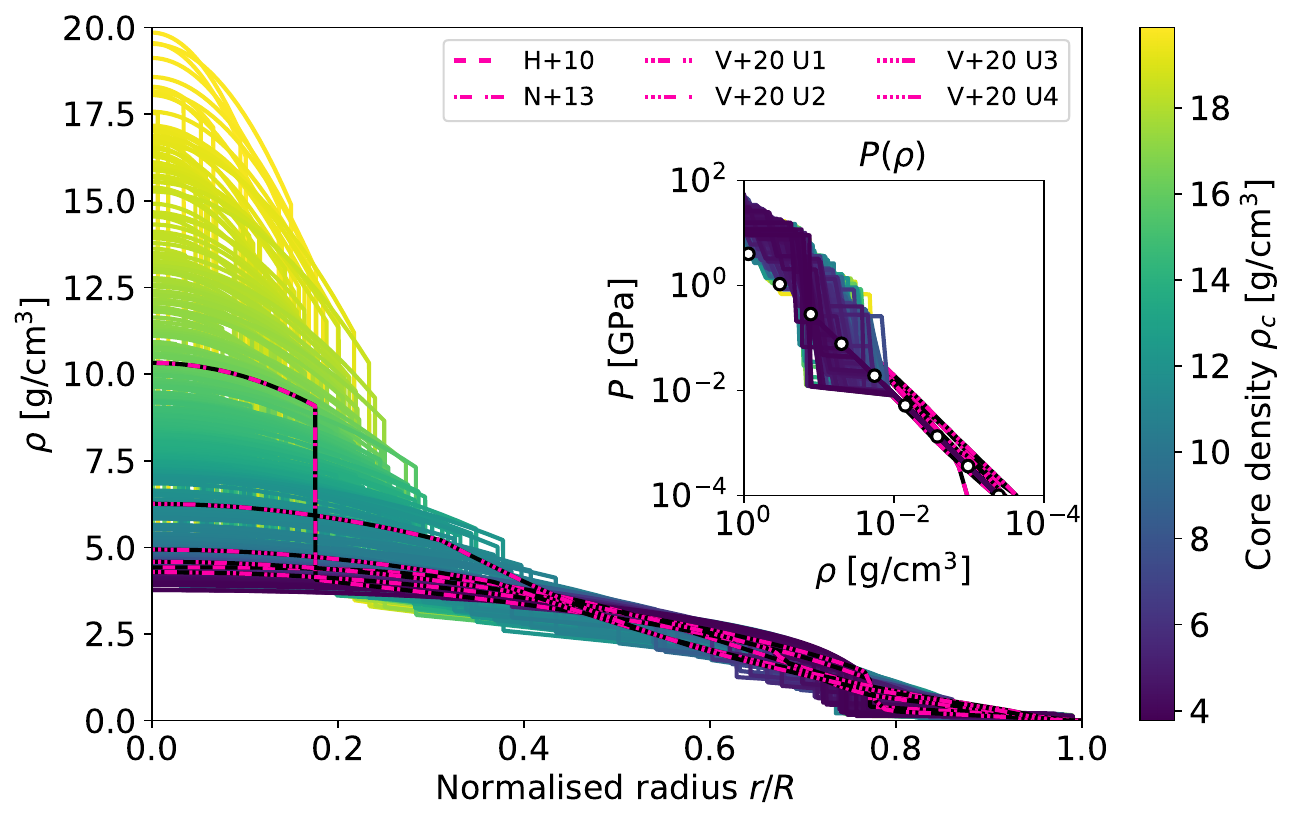}
    \caption{Uranus' density $\rho$ as a function of the average normalised  radius $r/R$.  Other density profiles that have been published by previous studies are also presented for comparison \citep{Helled_2010,Nettelmann_2013,Vazan_2020}. The panel on the right show the pressure as a function of density $P(\rho)$ in the outermost region. The white dots show the atmosphere model by \cite{Hueso_2020}. The full $P(\rho)$ distribution is presented in appendix \ref{app:data_summary} in Figure \ref{fig:P_of_rho}.}
    \label{fig:rho_of_r}
\end{figure}

Figure \ref{fig:rho_of_r} presents our 469 4-polytrope models for Uranus. The density $\rho$ is given as a function of the average normalised radius calculated by the ToF (see discussion below). We also show the pressure as a function of density $P(\rho)$ in the outermost part of the planet which fits an atmosphere model of Uranus \citep{Hueso_2020}. The atmosphere model is shown by the white dots and is followed by our models until $P(l^{*}) \geq P_{\mathrm{min}} = 10^{7}$ Pa. 

Overall, our model consists of four polytropes separated by three transition radii where only three polytropes are chosen freely to fit the gravitational moments $J_{2}$ and $J_{4}$. The measured gravitational moments include the non-hydrostatic contributions from Uranus' atmosphere dynamics \citep[e.g][]{Kaspi_2013,Soyuer_2020}. Our models assume uniform rotation and correspond to a fully hydrostatic planet without differential rotation or winds.  

\subsection{Tenth order Theory of Figures}
\label{sec:Method_ToF}

The models presented in Figure \ref{fig:rho_of_r} are based on the Theory of Figures (ToF) by \cite{Zharkov_1978}. The ToF inserts a spheroid Ansatz (surfaces parameterised by a function $r_{l}(\vartheta)$) into the total potential $U$. $\vartheta$ and $l$ are the polar angle and volumetric mean radius of the spheroid, respectively. $U = V + Q$ is the sum of the Newtonian potential $V = -G \int d^{3}r^{\prime} \rho/|\boldsymbol{r^{\prime}}-\boldsymbol{r}|$ and the centrifugal potential $Q = -\frac{1}{2}\omega^{2}r^{2}\sin^{2}(\vartheta)$. The details of inserting $r_{l}(\vartheta)$ into $U=U(r)$ and expanding the resulting expression order by order are discussed in detail in appendix \ref{app:ToF}. Originally, ToF was developed up to third order \citep{Zharkov_1978} and later was expanded to fourth \citep{Nettelmann_2017} and seventh order \citep{Nettelmann_2021}. For this work, we developed a tenth order ToF where the necessary expressions denoted as $f_{n}, f^{\prime}_{n}$, and $A_{n}$, which are themselves functions of so-called figure functions $s_{n}$, reach tenth order. Further details on our tenth order ToF implementation can be found in appendix \ref{app:ToF}. In subsection \ref{app:ToF_DiffRot}, we discuss potentials $Q$ that correspond to barotropic differential rotation and demonstrate that our ToF implementation accurately incorporates barotropic differential rotation.
\par
ToF calculates and converges on the figure and the gravitational moments $J_{2n}$ of the planet with a density profile $\rho(\vec{r})$. The gravitational moments appear in the expansion of the external part of the Newtonian potential into spherical harmonics: 
\begin{equation}
    \label{eq:ToF3}
    V_{\mathrm{ext}}(r,\mu) = \frac{GM}{r}\left(1-\sum_{n=1}^{\infty}\left(\frac{R_{\mathrm{eq}}}{r}\right)^{2n}J_{2n}P_{2n}(\mu)\right),
\end{equation}
where $R_{\mathrm{eq}}$ is the equatorial radius and $\mu = \cos(\vartheta)$. 

The ToF assumes no contributions of odd $J_{2n+1}$ and no dependence on the azimuthal angle $\varphi$. In order to calculate the $J_{2n}$, the ToF divides the planet into a given number of spheroids $N$. A larger $N$ leads to more accurate solutions for the figure and the planetary gravitational moments. 
\par
\begin{figure}
    \centering
    \includegraphics[width=\hsize]{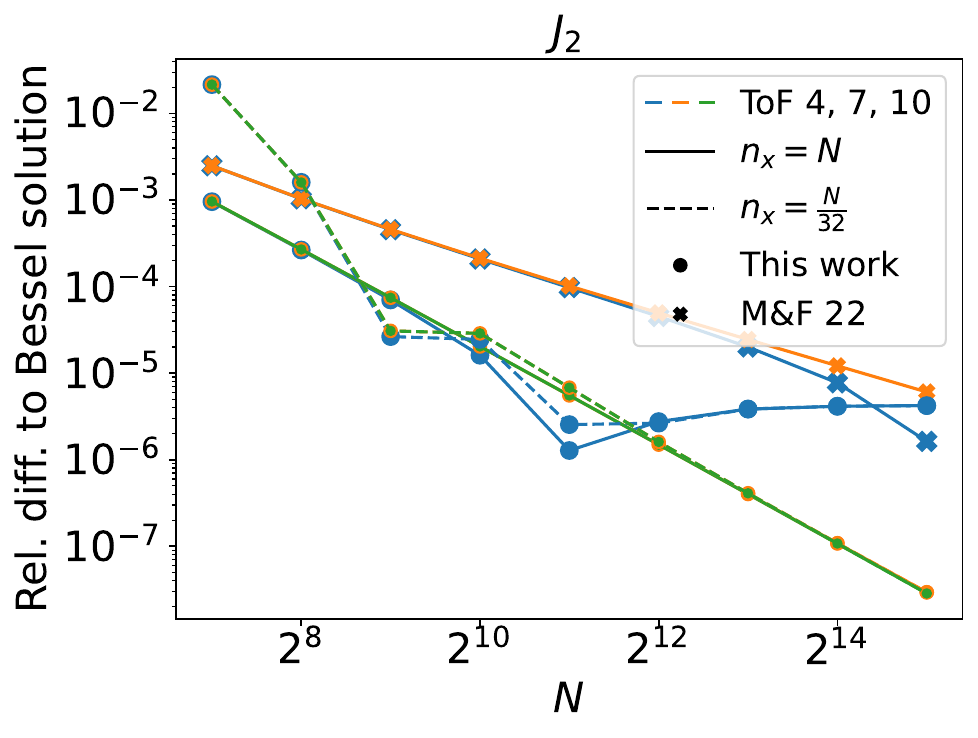}
    \includegraphics[width=\hsize]{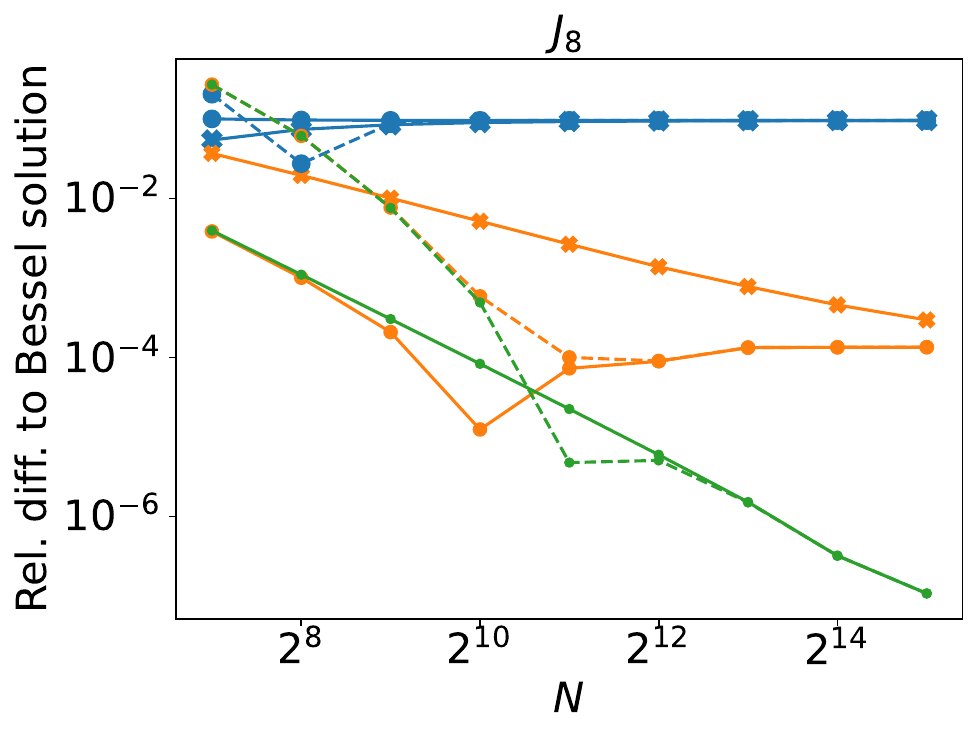}
    \caption{Relative difference of the Bessel gravitational moment solutions $J_2$ (top) and $J_8$ (bottom) compared to the ToF solutions. Results are shown for different numbers $N$ of used spheroids. We compare our implementation (This work) to the one used in  \cite{Movshovitz_2022} (M\&F 22) for different orders of the ToF. We also show the influence of using cubic spline interpolation with the parameter $n_x$ that stands for the number of points that were calculated without interpolation.}
    \label{fig:Bessel_convergence}
\end{figure}

In order to test the accuracy of our numerical ToF implementation, we use data provided by \cite{Wisdom_2016}. As a benchmark, they use a simple Jupiter model with a single polytropic Equation of State of $p = K \rho^2$ and assume a rotation period of 9$^{\text{h}}$55$^{\text{m}}$29.7$^{\text{s}}$ as well as an equatorial radius of $R_{\text{eq}} = 71492$ km. Thanks to the simplicity of this model, it is possible to independently calculate the gravitational moments $J_{2n}$ using spherical Bessel functions. We use these Bessel values of $J_{2n}$ as a reference and plot the relative difference to our solutions in Figure \ref{fig:Bessel_convergence} for $J_2$ (top) and $J_8$ (bottom) as examples. We present different orders of the ToF as well as the implementation used for \cite{Movshovitz_2022}. We also investigate the influence of approximating relevant expressions with spline interpolation. The parameter $n_x$ stands for the number of points that were calculated without interpolation. $n_x=N$ hence means that no interpolation was used, while $n_x=N/32$ means that only every 32nd point was actually calculated in the corresponding arrays, with all the intermediate points being estimated by using a cubic spline interpolation.

Several conclusions can be made from this comparison. 
First, our implementation of the ToF is more accurate than the one presented in \cite{Movshovitz_2022} and can reach relative precisions of the order of $10^{-7}$ instead of $10^{-3}$-$10^{-5}$ for the gravitational moments  up to $J_{8}$. Second, using seventh and tenth order ToF yield better accuracies for all the gravitational moments when using a sufficient high number of layers. Tenth order ToF is superior to seventh order ToF for gravitational moments beyond and including $J_{6}$. Third, the spline interpolation only influences accuracy when $N$ is smaller than $\sim2^{12}$.
\par
The implementation of the ToF presented here is more accurate than the one in \cite{Movshovitz_2022} and the two implementations have differences. 
The first difference between the two concerns a small typo in equation (B.17) for $f'_{4}$ in \cite{Nettelmann_2017}. The correct equation is\footnote{confirmed with Nadine Nettelmann in a private communication.}:

\begin{align}
    \label{eq:f4prime_typo}
    f'_4 &= \frac{1}{3}s_4 - \frac{9}{35}s_2^2 -\frac{20}{77}s_2 s_4 -\frac{45}{143} s_2 s_6 -\frac{81}{1001} s_4^2 \\
    &+ \frac{72}{385} s_2^3 + \frac{4579}{5005} s_2^2 s_4 - \frac{12798}{25025} s_2^4. \nonumber
\end{align}

Since \cite{Movshovitz_2022} relies on \cite{Nettelmann_2017} instead of equation \ref{eq:f4prime_typo}, they introduce a small error into their calculations for fourth order ToF. However, the resulting deviation is very marginal and only affects fourth order. This does not cause the differences visible in Figure \ref{fig:Bessel_convergence}. We believe that our improved accuracy is achieved by using the trapezoid rule to calculate the average density via $M = 4\pi\int \rho(r) r^2 dr$, instead of approximating $M \approx \sum_i \Delta\rho_i 4\pi r_i^3/3$\footnote{confirmed with Naor Movshovitz in a private communication.}.
\par
\begin{figure}[H]
    \centering
    \includegraphics[width=\hsize]{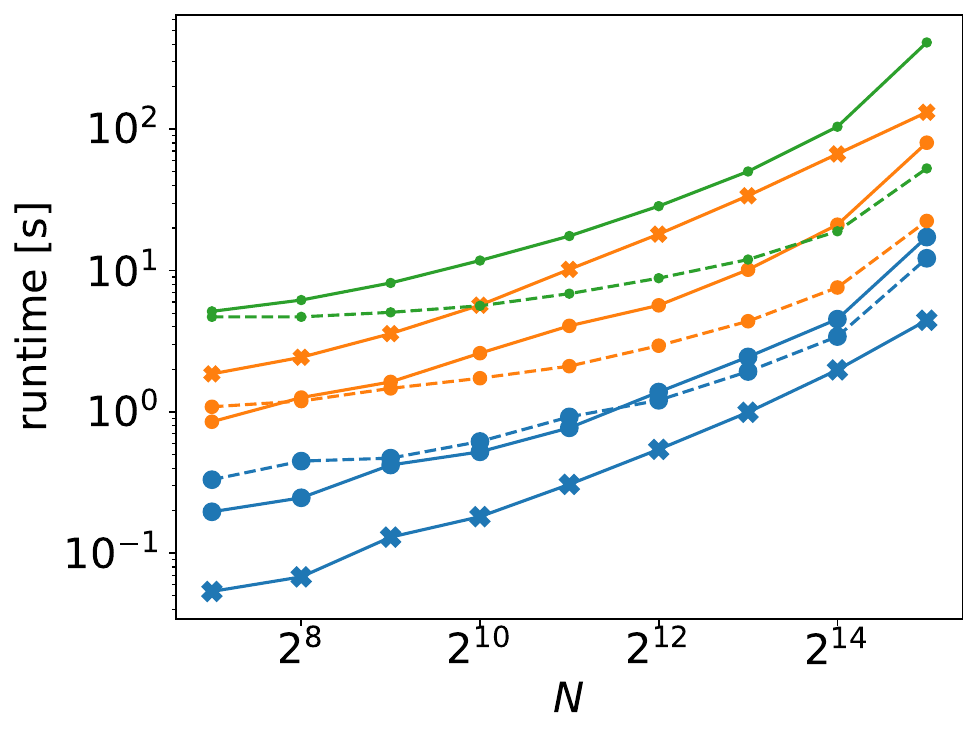}
    \caption{Runtime in seconds of the calculations shown in Figure \ref{fig:Bessel_convergence} as a function of the number $N$ of used spheroids in the ToF. We refer to the legend and caption of Figure \ref{fig:Bessel_convergence} for further explanations.}
    \label{fig:Bessel_time}
\end{figure}

Figure \ref{fig:Bessel_time} shows the runtime (in seconds) for the calculations depicted in Figure \ref{fig:Bessel_convergence}. The calculations were performed on a higher-end conventional laptop in Python. The ToF calculates all gravitational moments $J_{2n}$ simultaneously, so the presented runtime does not depend on a specific gravitational moment. 
We find that for fourth order ToF, the implementation of  \cite{Movshovitz_2022} is faster compared to our code. A spline interpolation does not lead to significant time savings. We also find that for higher orders, our implementation is faster. This has been achieved by directly computing the ToF coefficients in a machine-readable format and hence avoiding external coefficient files that need to be loaded repeatedly. Furthermore, we find that a spline interpolation can save significant computing time.

\tikzstyle{terminator} = [rectangle, draw, text centered, rounded corners, minimum height=2em]
\tikzstyle{process} = [rectangle, draw, text centered, minimum height=2em]
\tikzstyle{decision} = [diamond, draw, text centered, minimum height=2em]
\tikzstyle{connector} = [draw, -latex']
\tikzstyle{line} = [draw, -]

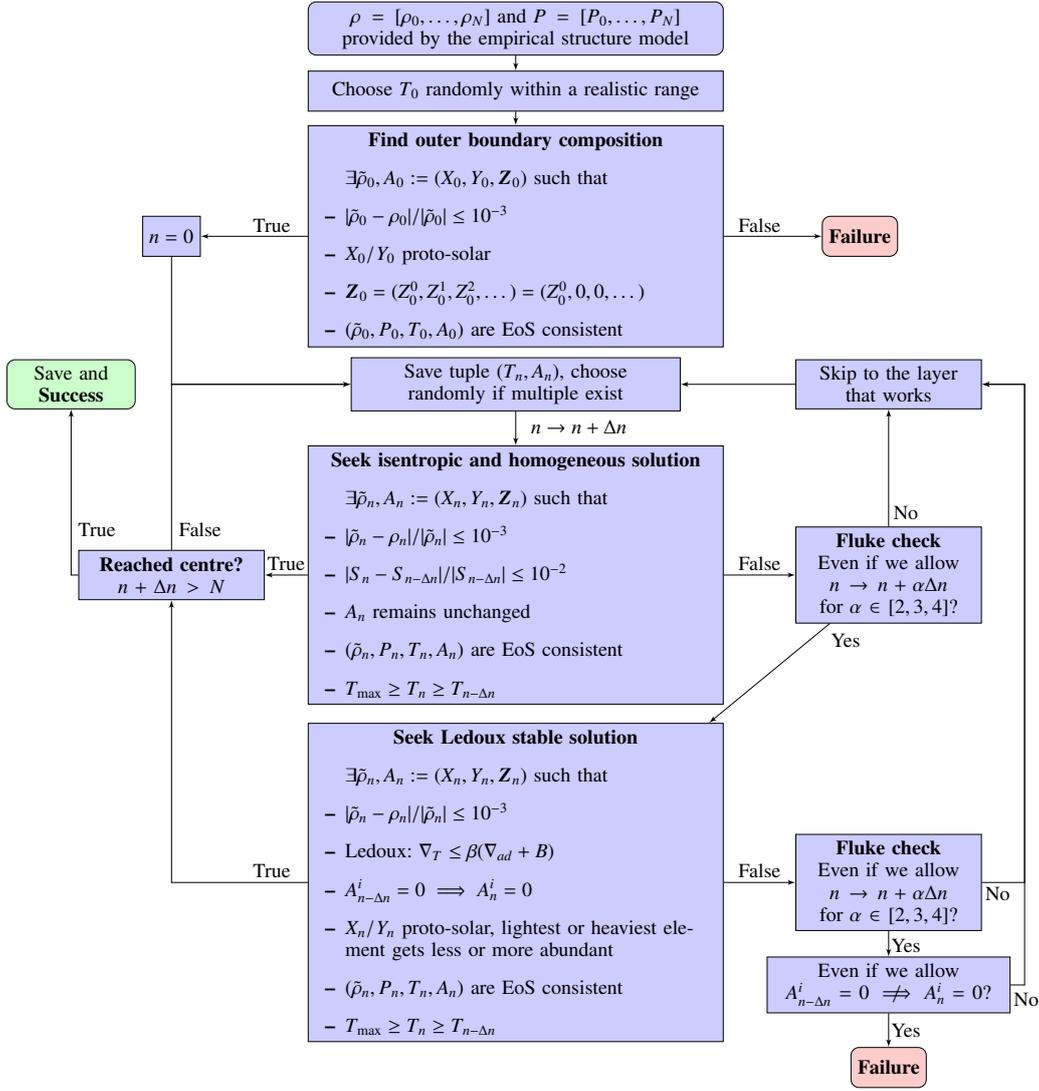
\begin{figure*}
\centering
\resizebox{0.75\hsize}{!}{

\begin{tikzpicture}
\node [terminator, fill=blue!20, text width=7cm] at (0,0) (start) {$\rho = [\rho_{0}, \dots, \rho_{N}]$ and $P = [P_{0}, \dots, P_{N}]$ provided by the empirical structure model};
\node [process, fill=blue!20, text width=7cm] at (0,-1.1) (choice1) {Choose $T_{0}$ randomly within a realistic range};
\node [process, fill=blue!20, text width=7cm] at (0,-3.65) (choice2) {
\textbf{Find outer boundary composition}
\begin{itemize}
    \item[] $\exists \Tilde{\rho}_{0}, A_{0} := (X_{0}, Y_{0}, \vec{Z}_{0})$ such that
    \item $|\Tilde{\rho}_{0}-\rho_{0}|/|\Tilde{\rho}_{0}| \leq 10^{-3} $
    \item $X_{0}/Y_{0}$ proto-solar
    \item $\vec{Z}_{0} = (Z_{0}^{0}, Z_{0}^{1}, Z_{0}^{2}, \dots) = (Z_{0}^{0}, 0, 0, \dots)$
    \item $(\Tilde{\rho}_{0}, P_{0}, T_{0}, A_{0})$ are EoS consistent
\end{itemize}
};
\node [terminator, fill=red!20] at (6,-3.65) (failure1) {\textbf{Failure}};
\node [process, fill=blue!20] at (-6,-3.65) (choice3) {$n=0$};

\node [process, fill=blue!20, text width=5.5cm] at (0,-6.25) (choice4) {Save tuple $(T_{n}, A_{n})$, choose randomly if multiple exist};
\node [process, fill=blue!20, text width=7cm] at (0, -9.6) (choice5) {
\textbf{Seek isentropic and homogeneous solution}
\begin{itemize}
    \item[] $\exists \Tilde{\rho}_{n}, A_{n} := (X_{n}, Y_{n}, \vec{Z}_{n})$ such that
    \item $|\Tilde{\rho}_{n}-\rho_{n}|/|\Tilde{\rho}_{n}| \leq 10^{-3} $
    \item $|S_{n}-S_{n-\Delta n}|/|S_{n-\Delta n}| \leq 10^{-2} $
    \item $A_{n}$ remains unchanged
    \item $(\Tilde{\rho}_{n}, P_{n}, T_{n},A_{n})$ are EoS consistent
    \item $T_{\mathrm{max}} \geq T_{n} \geq T_{n-\Delta n}$
\end{itemize}
};
\node [process, fill=blue!20, text width=3cm] at (-6, -9.6) (choice6) {
\textbf{Reached centre?}
$n + \Delta n > N$
};
\node [terminator, fill=green!20, text width=2cm] at (-7.75,-6.25) (success) {Save and \textbf{Success}};
\node [process, fill=blue!20, text width=3cm] at (6.5,-9.6) (choice7) {
\textbf{Fluke check} \\
Even if we allow $n \rightarrow n + \alpha \Delta n$ for $\alpha \in [2,3,4]$?};
\node [process, fill=blue!20, text width=3cm] at (6.5,-6.25) (choice8) {Skip to the layer that works};

\node [process, fill=blue!20, text width=7cm] at (0, -15) (choice9) {
\textbf{Seek Ledoux stable solution}
\begin{itemize}
    \item[] $\exists \Tilde{\rho}_{n}, A_{n} := (X_{n}, Y_{n}, \vec{Z}_{n})$ such that
    \item $|\Tilde{\rho}_{n}-\rho_{n}|/|\Tilde{\rho}_{n}| \leq 10^{-3} $
    \item Ledoux: $\nabla_T \leq \beta (\nabla_{ad} + B)$
    \item $A_{n-\Delta n}^{i} = 0 \implies A_{n}^{i} = 0$
    \item $X_{n}/Y_{n}$ proto-solar, lightest or heaviest element gets less or more abundant
    \item $(\Tilde{\rho}_{n}, P_{n}, T_{n},A_{n})$ are EoS consistent
    \item $T_{\mathrm{max}} \geq T_{n} \geq T_{n-\Delta n}$
\end{itemize}
};
\node [process, fill=blue!20, text width=3cm] at (6.5,-15) (choice10) {
\textbf{Fluke check} \\ 
Even if we allow $n \rightarrow n + \alpha \Delta n$ for $\alpha \in [2,3,4]$?};
\node [process, fill=blue!20, text width=4cm] at (6.5,-16.8) (choice11) {Even if we allow $A_{n-\Delta n}^{i} = 0 \centernot\implies A_{n}^{i} = 0$?};
\node [terminator, fill=red!20] at (6.5,-18.25) (failure2) {\textbf{Failure}};

\node[draw=none] at (4.25, -3.45) (false1) {False};
\node[draw=none] at (-4.25, -3.45) (true1) {True};
\node[draw=none] at (1.1, -7) (step) {$n \rightarrow n + \Delta n$};
\node[draw=none] at (-4, -9.4) (true2) {True};
\node[draw=none] at (-5.5, -8.8) (false2) {False};
\node[draw=none] at (-7.3, -8.8) (true3) {True};
\node[draw=none] at (4.25, -9.4) (false3) {False};
\node[draw=none] at (6.8, -8.5) (no1) {No};
\node[draw=none] at (5.75, -10.75) (yes1) {Yes};
\node[draw=none] at (4.25, -14.8) (false4) {False};
\node[draw=none] at (-4.25, -14.8) (true4) {True};
\node[draw=none] at (8.4, -15.2) (no2) {No};
\node[draw=none] at (6.8, -16.1) (yes2) {Yes};
\node[draw=none] at (8.9, -17.05) (no3) {No};
\node[draw=none] at (6.8, -17.6) (yes3) {Yes};

\node[draw=none] at (9,-15) (helper1) {};
\node[draw=none] at (8.87,-15.13) (helper2) {};

\node[draw=none] at (9,-16.8) (helper3) {};
\node[draw=none] at (8.87,-16.93) (helper4) {};

\path [connector] (start) -- (choice1);
\path [connector] (choice1) -- (choice2);
\path [connector] (choice2) -- (failure1);
\path [connector] (choice2) -- (choice3);
\path [connector] (choice3) |- (choice4);
\path [connector] (choice4) -- (choice5);
\path [connector] (choice5) -- (choice6);
\path [connector] (choice6) -| (success);
\path [connector] (choice6) |- (choice4);
\path [connector] (choice5) -- (choice7);
\path [connector] (choice7) -- (choice8);
\path [connector] (choice8) -- (choice4);
\path [connector] (choice7) -- (choice9);
\path [connector] (choice9) -- (choice10);
\path [line] (choice10) -- (helper1);
\path [connector] (helper2) |- (choice8);
\path [connector] (choice10) -- (choice11);
\path [connector] (choice11) -- (failure2);
\path [connector] (choice9) -| (choice6);
\path [line] (choice11) -- (helper3);
\path [connector] (helper4) |- (choice8);

\end{tikzpicture}

}

\caption{Our algorithm for inferring the temperature profile and composition of the empirical density (and pressure) profiles.}
\label{fig:Algorithmus}
\end{figure*}

\subsection{A new algorithm to infer the temperature profile and composition} 
\label{sec:Method_Algorithm}

In this subsection we describe the new algorithm to infer the composition and temperature $T$ profiles  for the empirical models. We improve over previous work \citep[for example][]{Podolak_2022,Neuenschwander_2024} by introducing randomness into the process. This allows us to find a significantly wider range of possible solutions for each density profile, which is more consistent with the spirit of empirical models to find 'outside the box' solutions. The main steps of the algorithm can be summarised as follows:

\begin{itemize}
    \item We begin with an empirical structure model that provides the planetary density $\rho = [\rho_{0}, \dots, \rho_{N}]$ and pressure $P = [P_{0}, \dots, P_{N}]$. Since our EoS for Hydrogen and Helium (H-He) are limited to temperatures above 100 K \citep{Chabrier_2019}, the first few entries of $\rho$ and $P$ are  skipped.
    \item The algorithm tries to find a starting tuple $(\rho_{0}, P_{0}, T_{0}, X_{0}, Y_{0}, \vec{Z}_{0})$ that is consistent with the EoS and the empirical profile and chooses randomly if multiple exist.
    \item The algorithm continues to the next spheroid (henceforth called layers) with a stepsize $\Delta n$, which we took to be 50 or 100 for our $N = 4096$ models. Smaller step sizes only lead to larger computing times with no significant change to our results. 
    \item The algorithm tries to find isentropic solutions first. With this choice, we bias our algorithm towards finding isentropic solutions (corresponding to convective regions). 
    \item If there is no isentropic solution, we try changing the composition. Priority is given to changing the composition of the materials already present. If that fails,  heavier components are added. In both cases, we check the Ledoux criterion introduced below and choose randomly if multiple options exist.
    \item This is repeated until the planetary centre is reached or it is impossible to generate new candidates $(\rho_{n}, P_{n}, T_{n}, X_{n}, Y_{n}, \vec{Z}_{n})$ that fulfil all necessary conditions.
\end{itemize}

The details of our algorithm are summarised in the flow chart presented Figure \ref{fig:Algorithmus}. We allow the skipping and ignoring of up to three layers to ensure that changing the composition is truly necessary and not just a fluke due to the uncertainties of the EoS and the empirical model. Due to the uncertainties of the EoS, we set $|S_{n}-S_{n-\Delta n}|/|S_{n-\Delta n}| \leq 10^{-2}$ for the isentropic condition.
\par
Our implementation of the random algorithm in Figure \ref{fig:Algorithmus} supports up to four materials at the same time, using the assumption of ideal mixing for the total density $\rho$:
\begin{equation}
    \label{eq:ideal_mixing}
    \frac{1}{\rho} = \frac{X+Y}{\rho_{X+Y}} + \frac{Z^{0}}{\rho_{Z^{0}}} + \frac{Z^{1}}{\rho_{Z^{1}}} + \frac{Z^{2}}{\rho_{Z^{2}}}.
\end{equation}
The ratio of the Hydrogen and Helium (H-He) mass fractions $X$ and $Y$ are fixed to be proto-solar. Therefore, they are treated as a single component. As a result, our algorithm can treat up to four materials simultaneously, distributed into three (composite) components. We use the \cite{Chabrier_2019} EoSs for the H-He mixture. For the heavier materials $\vec{Z}$ we assume water (H$_{2}$O), rocks (SiO$_{2}$), and iron (Fe) using the EoSs of \cite{More_1988}. 
\par
Often, the interiors of the giant planets in the Solar System are assumed to be isentropic due to large-scale convection. 
However, non-isentropic regions can exist and could be a result of boundary layers or composition gradients. Both need to be locally stable, which is tested by displacing a bubble of material. If the bubble oscillates around its original position (with a frequency $N \in \mathbb{R}$), the surrounding is locally stable. If the frequency $N \in i\mathbb{R}$ is imaginary, the bubble gets pushed further away from its original position. The surrounding is interpreted to be unstable, that is, subjected to convection. To check for local stability in our algorithm, we use the slightly modified Ledoux criterion from \cite{Paxton_2013}. Following \cite{Kippenhahn_2013}, the Brunt-Väisälä frequency $N$ of a displaced bubble of material is \citep[see also][for details]{Unno_1989}:

\begin{equation}
    \label{eq:N_squared}
    N^{2} = \frac{g^{2}\rho}{P}\frac{\chi_{T}}{\chi_{\rho}}\left[ \nabla_{ad} - \nabla_T +\underbrace{\frac{\chi_{\rho}}{\chi_{T}}\left(\frac{\partial \ln \rho}{\partial \ln \mu}\right)_{P,T}\frac{\mathrm{d}\ln\mu}{\mathrm{d}\ln\rho}\frac{\mathrm{d}\ln\rho}{\mathrm{d}\ln P}}_{=:B}\right].
\end{equation}

Following similar steps as in \cite{Paxton_2013}, $B$ can be calculated numerically as: 

\begin{equation}
    \label{eq:brunt_B}
    B = \frac{\chi_{\rho}}{\chi_{T}}\frac{\ln \rho \left(P_{n}, T_{n}, \vec{A}_{n+\Delta n}, \right) - \ln \rho \left(P_{n}, T_{n}, A_{n} \right)}{\ln P_{n+\Delta n} - \ln P_{n}},
\end{equation}

where $\vec{A}_{n} := (X_{n}, Y_{n}, \vec{Z}_{n})$. We provide a detailed derivation in appendix \ref{app:Ledoux}. Requiring $N^2 \geq 0$ leads to the Ledoux criterion stated in Figure \ref{fig:Algorithmus}: 

\begin{equation}
    \nabla_T \leq \beta (\nabla_{ad} + B),
\end{equation}

where 

\begin{equation}
    \nabla_{ad} := \left( \frac{\partial \ln T}{\partial \ln P} \right)_{S}, \quad \nabla_T := \frac{d \ln T}{d \ln P}.
\end{equation}

The $\beta$ pre-factor $\in [0,1]$ assists the algorithm to find lower central temperature solutions more quickly. Excluded solutions with $\beta (\nabla_{ad} + B) \leq \nabla_T < \nabla_{ad} + B$ would have higher temperature gradients. We routinely used $\beta=0.7$ and still found plenty of high central temperature solutions (see Section \ref{sec:Results}). 
\par
The $\beta$ pre-factor parameterises the uncertainty of the energy transport in inhomogeneous layers. If the composition gradient is stable there are two possibilities for the energy transport. The first is double-diffusive convection and the second is conduction. For planetary interior conditions, double-diffusive convection is still poorly understood. It is unclear whether layers can form and survive on long timescales \citep{Wood_2013,Fuentes_2022}. In addition, it is unclear what the appropriate temperature profile is when composition gradients exist \citep[for example][]{Garaud_2018}. For thermal conduction, the temperature gradient is well-defined in principle. However, it depends on an unknown local luminosity in the deep interior and the thermal conductivity, which is highly uncertain for the relevant conditions. Given these issues, we remain agnostic about the temperature gradient and let the algorithm randomly pick from a temperature consistent with the empirical model and Ledoux stability. Our choice of $\beta = 0.7$ is motivated by the fact that it is unlikely that the temperature gradient is very similar to the Ledoux gradient. Such regions would only be marginally stable and could also lead to extreme temperatures. 

\begin{figure}

    \centering
    \includegraphics[width=\hsize]{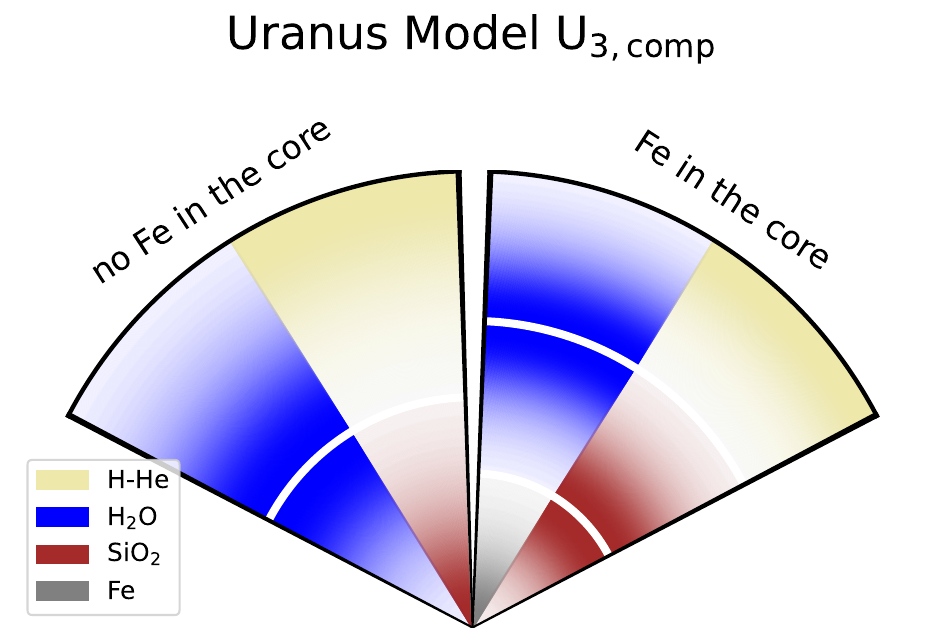}
    \includegraphics[width=\hsize]{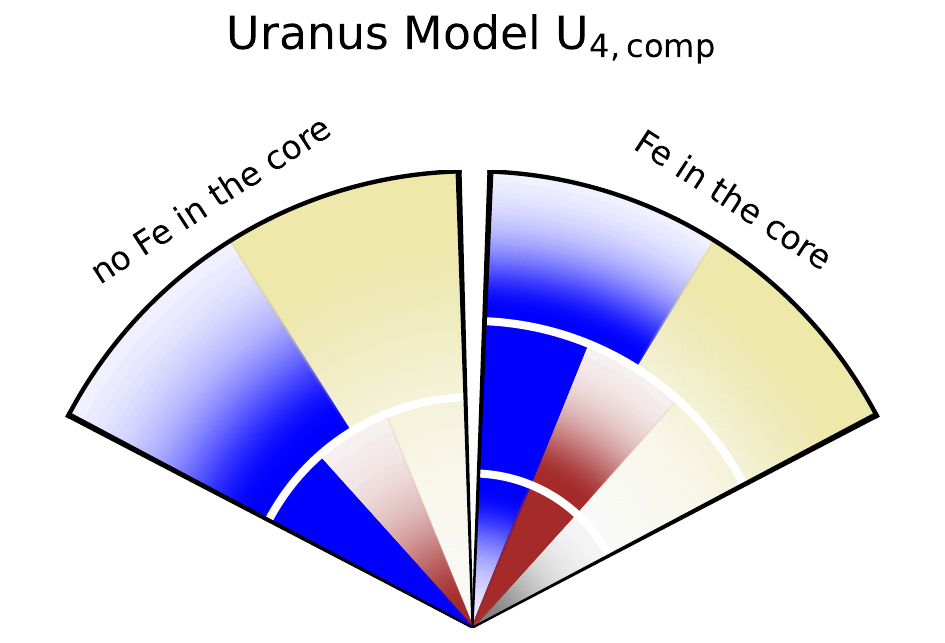}

    \caption{Schematic drawing of the two possible composition models considered in this work. Top: U$_{3,\text{comp}}$ model that allows the presence of up to three elements simultaneously in a given layer. Bottom: U$_{4,\text{comp}}$ model that allows the presence of up to four elements simultaneously in a given layer. The ratios between different components are chosen freely and composition gradients are allowed but not necessary. The transition radii (white gaps) vary and are placed by the algorithm.}
    \label{fig:3UM4UM}
    
\end{figure}

Figure \ref{fig:3UM4UM} presents the two different interior models we consider in  this study to infer Uranus' composition. The top panel corresponds to a model of Uranus admitting up to three materials   per layer (U$_{3,\text{comp}}$). We begin with H-He and H$_{2}$O in Uranus' atmosphere. As soon as the algorithm fails to find more solutions with these components, we allow the addition of SiO$_{2}$ and remove H-He. When no solutions can be found with H$_{2}$O and SiO$_{2}$, then H$_{2}$O is removed and we allow for the addition of Fe. This model is more traditional in the sense that it follows a structure in which lighter  elements do not exist in the deep interior. The bottom panel on the other hand shows a model of Uranus admitting up to four materials per layer (U$_{4,\text{comp}}$). Just as before, we begin with H-He and H$_{2}$O in the outermost part of the planet. As soon as the algorithm fails to find more solutions with these components, we allow the addition of SiO$_{2}$, but still allow for the presence of H-He. If at some point  no  solutions with H-He, H$_{2}$O, and SiO$_{2}$ can be found, then H-He are removed and we allow for the addition of Fe. This configuration allows solutions with 'fuzzy cores' as we allow H-He to exist in Uranus' deep interior \citep{Helled_2017,Valletta_2022}. 

\section{Results}
\label{sec:Results}

In total, the random algorithm derived composition-temperature profiles for 469 empirical density profiles (shown in Figure \ref{fig:rho_of_r}), which yielded 14659 U$_{4,\text{comp}}$ and 13645 U$_{3,\text{comp}}$ solutions. First, we present our predictions for the gravitational moments beyond $J_{6}$ up to $J_{14}$ using our updated empirical models (Figure \ref{fig:rho_of_r}) in Subsection \ref{sec:Results_Jvalues}. In Subsection \ref{sec:Results_Profiles} we present the inferred composition and temperature profiles using the new algorithm for three examples of our U$_{4,\text{comp}}$ inferred models. We compare our complete results for the two cases U$_{3,\text{comp}}$ and U$_{4,\text{comp}}$ in subsection \ref{sec:Results_3vs4} and discuss relevant correlations between parameters of interest in subsection \ref{sec:Results_Correlations}.

\subsection{Gravitational moments predictions}
\label{sec:Results_Jvalues}

Our tenth order ToF allows us to calculate Uranus' gravitational moments accurately and to high order. We first use our models to predict Uranus' gravitational coefficients $J_{6}$ up to $J_{14}$ presented in Table \ref{tab:res_Jvalues}. We find that $J_{6} = (5.3078 \pm 0.3312) \cdot 10^{-7}$ and $J_{8} = (-1.1114 \pm 0.1391) \cdot 10^{-8}$. These values are in agreement with the values predicted by \cite{Neuenschwander_2022}. For the higher order gravitational moments, we find $J_{10} = (2.8616 \pm 0.5466) \cdot 10^{-10}$, $J_{12} = (-8.4684 \pm 2.0889) \cdot 10^{-12}$ and $J_{14} = (2.7508 \pm 0.7944) \cdot 10^{-13}$. These values could be measured via a future Uranus mission and can constrain Uranus' internal structure \citep[for example][]{Helled_2020_1} and atmosphere dynamics \citep[for example][]{Kaspi_2017, Soyuer_2020}.

\begin{table}[H]
\centering       
\caption{Predictions of Uranus' gravitational moments ($J_{6}-J_{14}$) for the 4-polytrope empirical models presented in figure \ref{fig:rho_of_r} assuming uniform rotation. The ranges indicate the 1$\sigma$ region.} 
\begin{tabular}{l|l}
\hline
$J_{6}$  & $ (\phantom{+} 5.3078 \pm 0.3312) \cdot 10^{-7} $ \\
$J_{8}$  & $ (         -  1.1114 \pm 0.1391) \cdot 10^{-8} $ \\
$J_{10}$ & $ (\phantom{+} 2.8616 \pm 0.5466) \cdot 10^{-10}$ \\
$J_{12}$ & $ (         -  8.4684 \pm 2.0889) \cdot 10^{-12}$ \\
$J_{14}$ & $ (\phantom{+} 2.7508 \pm 0.7944) \cdot 10^{-13}$ \\
\hline
\end{tabular}%

\label{tab:res_Jvalues} 
\end{table}

\subsection{Composition and temperature profiles}
\label{sec:Results_Profiles}

\begin{figure*}
\centering
\resizebox{\hsize}{!}{
    \centering
    \begin{subfigure}[b]{0.33\textwidth}
        \centering
        \includegraphics[width=\textwidth]{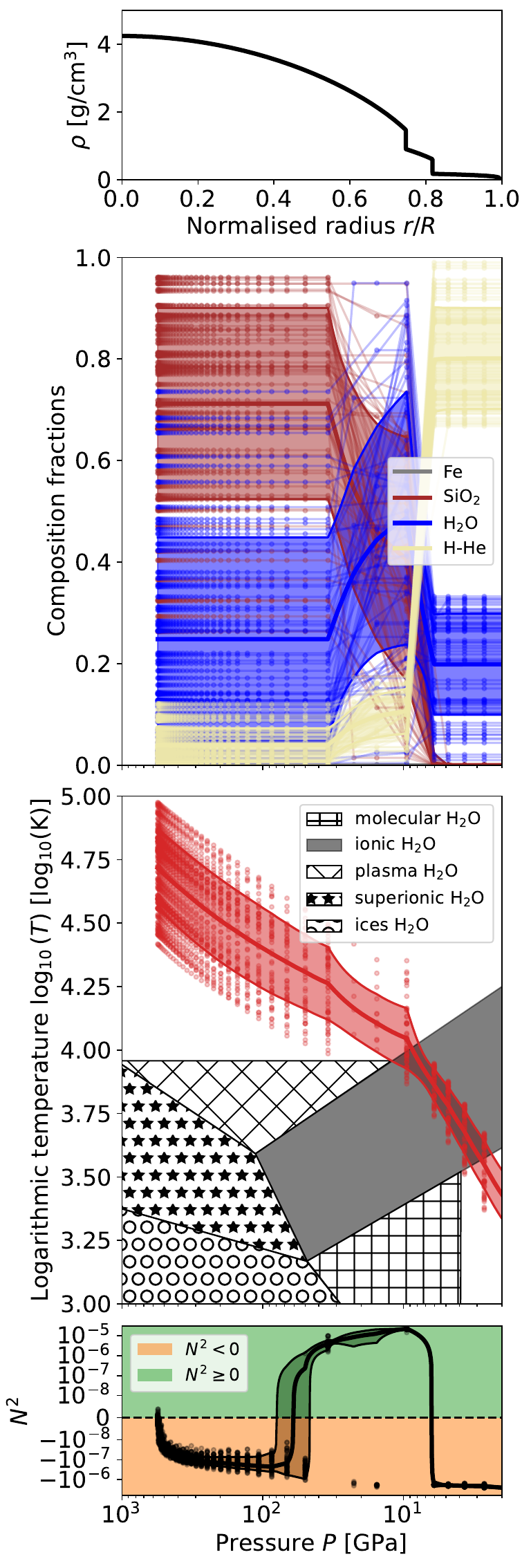}
        \caption{Low central density, mostly convective}
        \label{fig:SampleResults_A}
    \end{subfigure}
    \hfill
    \begin{subfigure}[b]{0.33\textwidth}
        \centering
        \includegraphics[width=\textwidth]{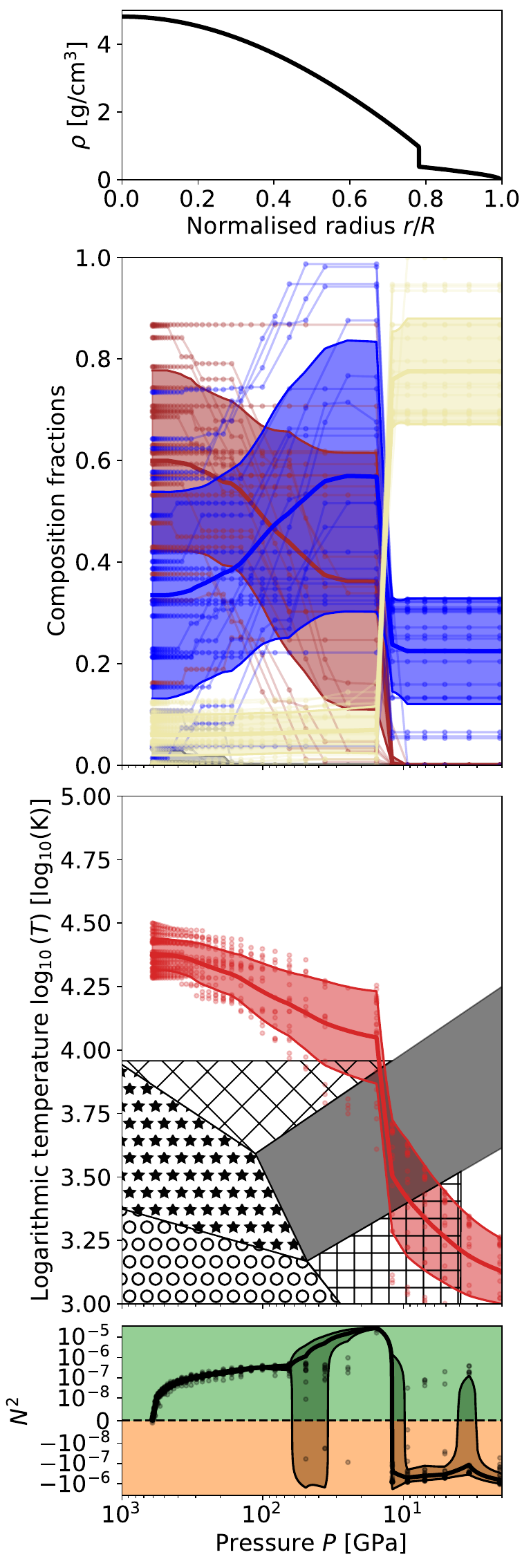}
        \caption{Low central density, mostly non-convective}
        \label{fig:SampleResults_B}
    \end{subfigure}
    \hfill
    \begin{subfigure}[b]{0.33\textwidth}
        \centering
        \includegraphics[width=\textwidth]{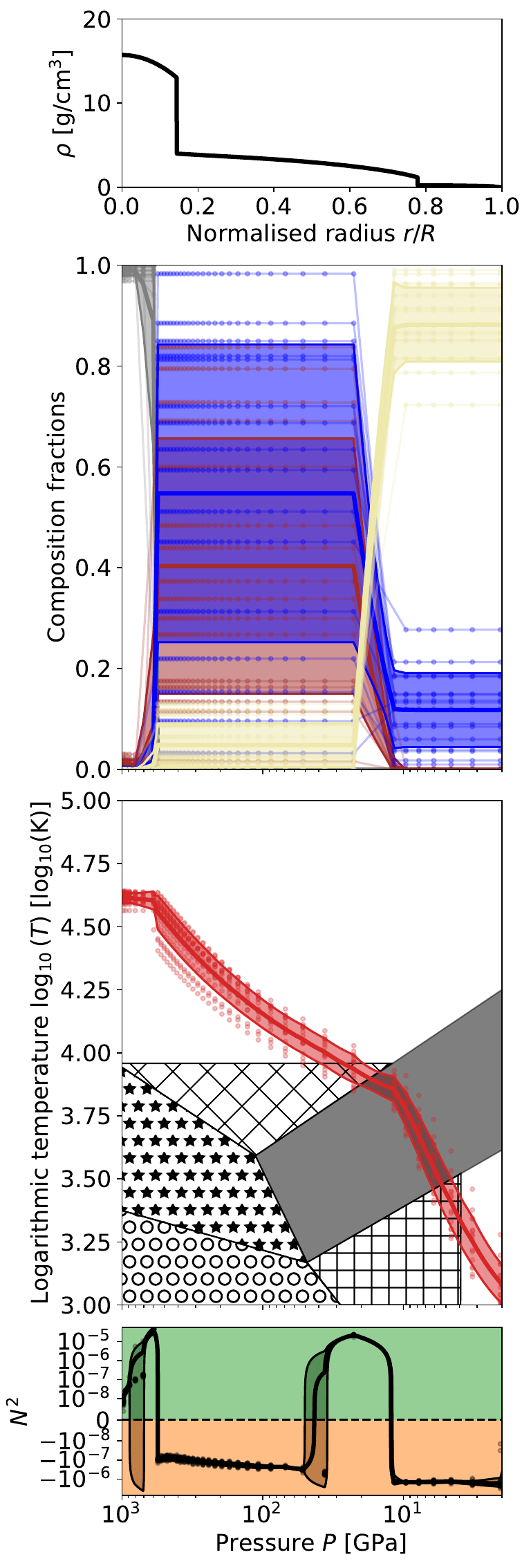}
        \caption{High central density, mostly convective}
        \label{fig:SampleResults_C}
    \end{subfigure}
    }
    \caption{Three example empirical density profiles of Uranus (depicted from left to right). Their density, composition, and temperature profile inferred by the new random algorithm is depicted from top to bottom, respectively. The extended phase diagram of water based on \cite{Redmer_2011} is shown in addition to the temperature profiles. The stability against convection is presented using the Brunt-Väisälä frequency $N^{2}$ in the bottom last plots.}
    \label{fig:SampleResults}
\end{figure*}

In this subsection we present the composition and temperature profiles and identify the stable regions against convection using the random algorithm. First, we show three different empirical density profiles (U$_{4,\text{comp}}$ model) which are presented in Figure \ref{fig:SampleResults}. These three models were chosen as examples to illustrate the diversity of possible solutions. The dotted lines correspond to individual solutions. The average of all individual solutions of a given profile is shown by a thick line and the 1$\sigma$ range is shown by the coloured region. 

The low central density profile presented in Figure \ref{fig:SampleResults_A} leads to solutions with a well-mixed and rock-rich inner region. We find that there are two large convection zones (with a corresponding Brunt-Väisälä frequency of $N^{2} < 0$) separated by a broad stable layer at $P \sim 20$ GPa and $\sim 0.75 r_{U}$, where  $r_{U}$ is the average surface radius of Uranus. H-He is present throughout the planet, even in the centre with small proportions. The inferred central temperatures are high, of the order of 50,000 K.
\par
On the other hand, the model presented in Figure \ref{fig:SampleResults_B} which also has a low central density and a somewhat similar density distribution, is found to have a rather different solution. This model is incompatible with a well-mixed inner region and is found to have only one large convection zone in the outer region of the planet that stops at $P \sim 20$ GPa and $\sim 0.75 r_{U}$. The deep interior consists of composition gradients including H-He with typical central temperatures of $\sim$20,000 K. Although at first sight the two density profiles seem rather similar they are sufficiently different to result in significantly different solutions for the inferred temperature and composition profiles. This clearly illustrates the problem of degeneracy in interior modelling. A direct comparison of the two density profiles is presented in appendix \ref{app:data_summary} in Figure \ref{fig:compare30vs160}. 
\par
Finally, the high central density profile shown in Figure \ref{fig:SampleResults_C} is characterised by three distinct zones. First, there are two zones that are clearly convective and are separated by a sharp stable layer at $\sim 0.8 r_{U}$ and $P \sim 10$ GPa. The water-to-rock ratio of the second middle zone varies significantly for different individual solutions. The final innermost zone is found to be iron dominated in composition, with only negligible amounts of rocks and water. H-He is present in the middle zone, but not in the planetary centre. The central temperatures are high and are  found to be $\sim$40,000 K on average. 
\par
All of these profiles possess a convective ionic water region (according to the $p$-$T$ line in \cite{Redmer_2011}), consistent with the requirement to generate Uranus' magnetic field. The three following criteria are fulfilled at the same time: $N^{2}<0$ holds, H$_{2}$O is present in the composition mix and the red temperature lines cross the ionic H$_{2}$O region. This could explain the origin of Uranus' magnetic field that can be attributed to the high electrical conductivity profile in this region \citep[for example][and references therein]{Soyuer_2020}. The majority (77\% in the U$_{3,\text{comp}}$ case and 83\% in the U$_{4,\text{comp}}$ case) of our profiles were inferred to have a convective ionic water region.
\par
The inferred magnitudes of the Brunt-Väisälä frequencies in convective regions are unrealistically large. This is likely because the polytropic models can have density gradients different than expected from an adiabatic and homogeneous region, which would be too unstable \citep[see also][]{Neuenschwander_2024}. Here we only examine whether $N^2$ is negative or positive to interpret our profiles regarding convective stability of the regions with composition gradients. In Section \ref{sec:discussion_temperature} we discuss the temperature-sensitivity of the inferred composition. Future work should address this topic and ensure to provide realistic density gradients and Brunt-Väisälä frequencies, which could provide further constraints when modelling Uranus' interior.

\subsection{U$_{3,\text{comp}}$ versus U$_{4,\text{comp}}$}
\label{sec:Results_3vs4}

Below, we present the results of all the empirical models combined. We begin with  comparing these two cases focusing on the inferred composition and temperature profiles. 

\subsubsection*{Uranus' bulk composition}

\begin{figure}

    \centering
    \includegraphics[width=\hsize]{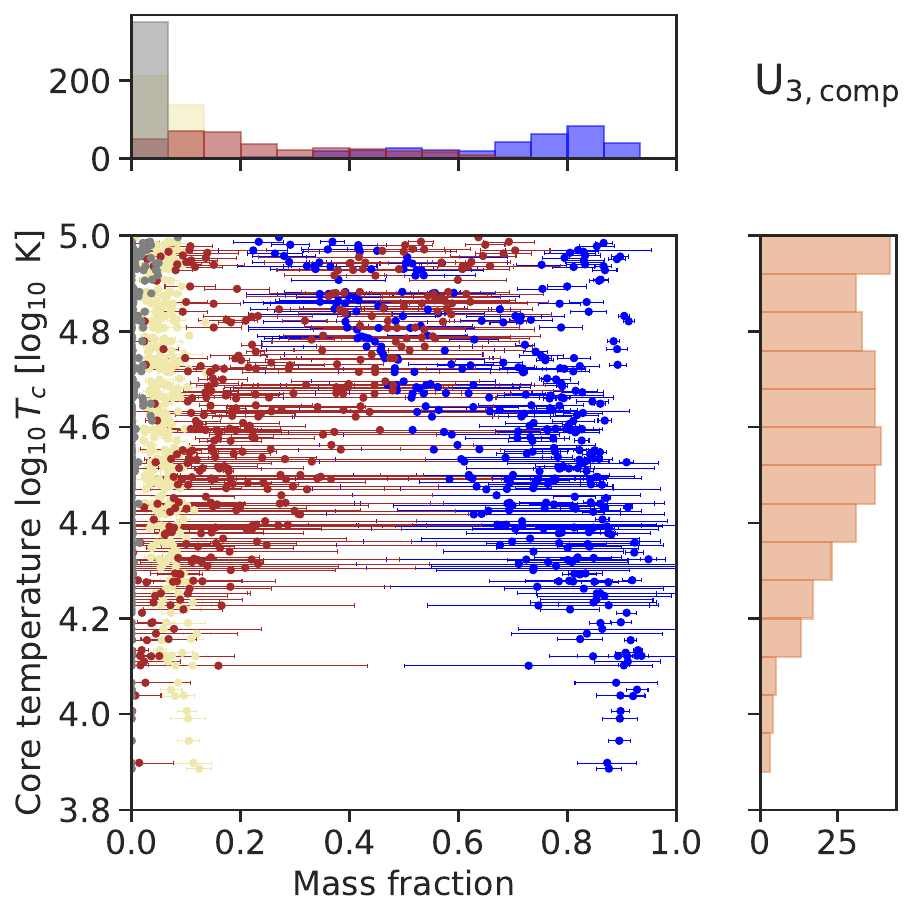}
    \includegraphics[width=\hsize]{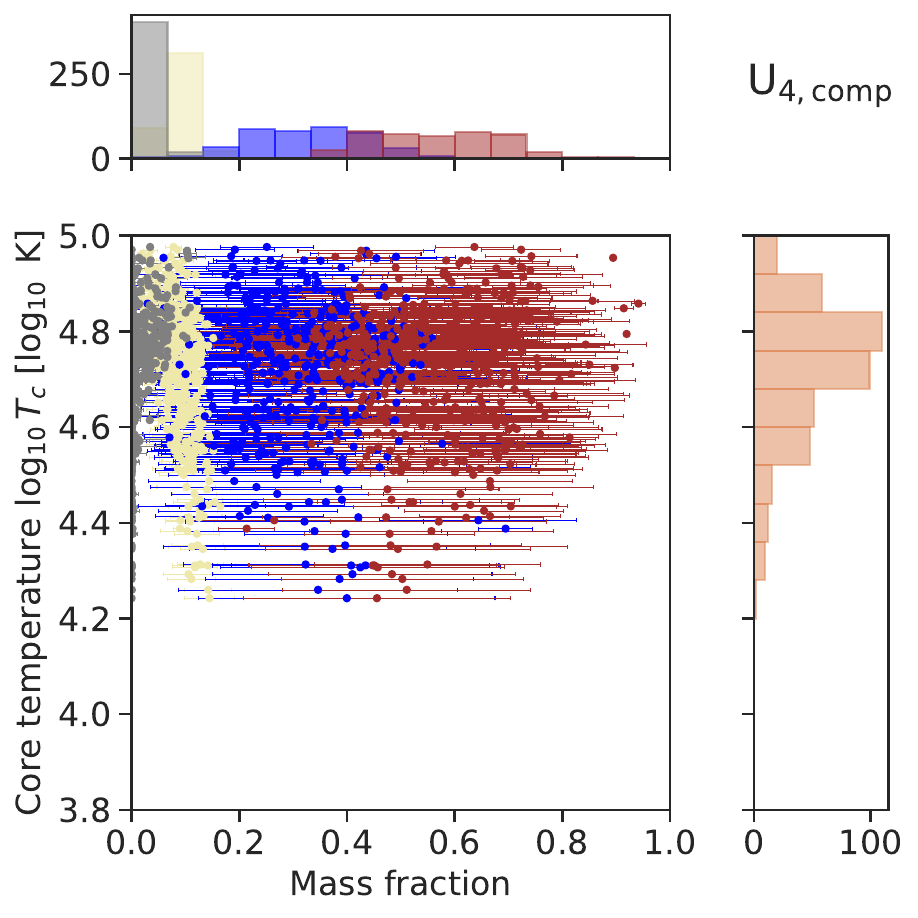}

    \caption{Inferred composition and temperature for our Uranus empirical models for U$_{3,\text{comp}}$ (top) and U$_{4,\text{comp}}$ (bottom). The x-axis corresponds to the mass fraction of a material and the y-axis shows the central temperature. The colours represent different materials (see legend of Figure \ref{fig:SampleResults}) and the error bars indicate the $1\sigma$ range of solutions found by the random algorithm for each density profile. To improve readability, error bars for the central temperature $T_{c}$ are not shown. They are included in Figure \ref{fig:corr_h_he_centre}.}
    \label{fig:Results_Compositions}
    
\end{figure}

The assumed internal structure significantly affects the inferred bulk composition. Figure \ref{fig:Results_Compositions} shows the complete range of inferred compositions and central temperatures for cases U$_{3,\text{comp}}$ (top) and U$_{4,\text{comp}}$ (bottom). 
We find that most of the solutions for the U$_{3,\text{comp}}$ model are water-rich and that rock-dominated interiors are rare. The few rock-dominated solutions we found require extreme central temperatures exceeding $\sim 50,000$ K that are rather unrealistic. Figure \ref{fig:Results_Compositions} clearly shows for the U$_{3,\text{comp}}$ case that the water mass fraction is inversely correlated with the central temperature, where lower temperatures correspond to more water-rich interiors. More specifically, the water mean mass fraction values possess a Spearman's rank correlation coefficient of -0.57 at a p-value
of < 0.005 when compared to the mean central temperatures. 

Unlike the U$_{3,\text{comp}}$ case, the U$_{4,\text{comp}}$ models allow H-He to exist in the deep interior. This significantly affects the inferred rock mass fraction and rock-dominated solutions are found to be much more common. Adding H-He to the mixture of materials compensates for higher abundances of heavier materials (such as SiO$_{2}$). This in turn influences the central temperatures, which are now always hotter than $\sim 15,000$ K (compared to $\gtrsim 8,000$ K for the U$_{3,\text{comp}}$ models). There is also no correlation between the water mass fraction and the central temperature. The water mean mass fraction values now possess a Spearman's rank correlation coefficient of -0.03 at a p-value of 0.55 when compared to the mean central temperatures. Furthermore, we find that typically water is less common than rock for all central temperatures. 
\par
\begin{figure}
    \centering
    \includegraphics[width=\hsize]{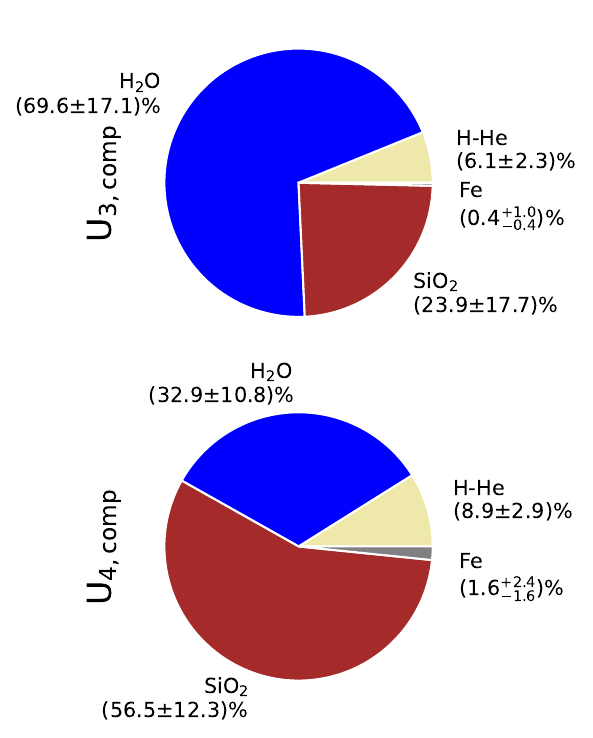}
    \caption{Uranus' inferred composition using the new random algorithm. We depict results for the U$_{3,\text{comp}}$ (top) and U$_{4,\text{comp}}$ (bottom) models. The pie charts summarise the composition histograms from Figure \ref{fig:Results_Compositions}, and the numbers are the means and 1$\sigma$ ranges of all empirical models shown as dots in Figure \ref{fig:Results_Compositions}.}
    \label{fig:Results_Compositions_Easy}
\end{figure}

We next summarise all composition results in Figure \ref{fig:Results_Compositions_Easy}. It shows the overall mass fraction of the different materials for both scenarios U$_{3,\text{comp}}$ (top) and U$_{4,\text{comp}}$ (bottom). The dots displayed in Figure \ref{fig:Results_Compositions} are weighted averages to account for the fact that we did not necessarily find an equal amount of composition-temperature profiles for each empirical density profile. Figure \ref{fig:Results_Compositions_Easy} presents the mean and 1$\sigma$ range of all these weighted averages. 
We find that for the U$_{3,\text{comp}}$ models water is generally the most abundant material, followed by rocks, H-He, and iron. However, from the U$_{4,\text{comp}}$ models, it is clear that if H-He can exist in the deeper interior we generally expect a water-to-rock ratio that tends to be significantly smaller than one ($\sim 0.6$).

\subsubsection*{Convectivity of Uranus' interior}

\begin{figure}

    \centering
    \includegraphics[width=\hsize]{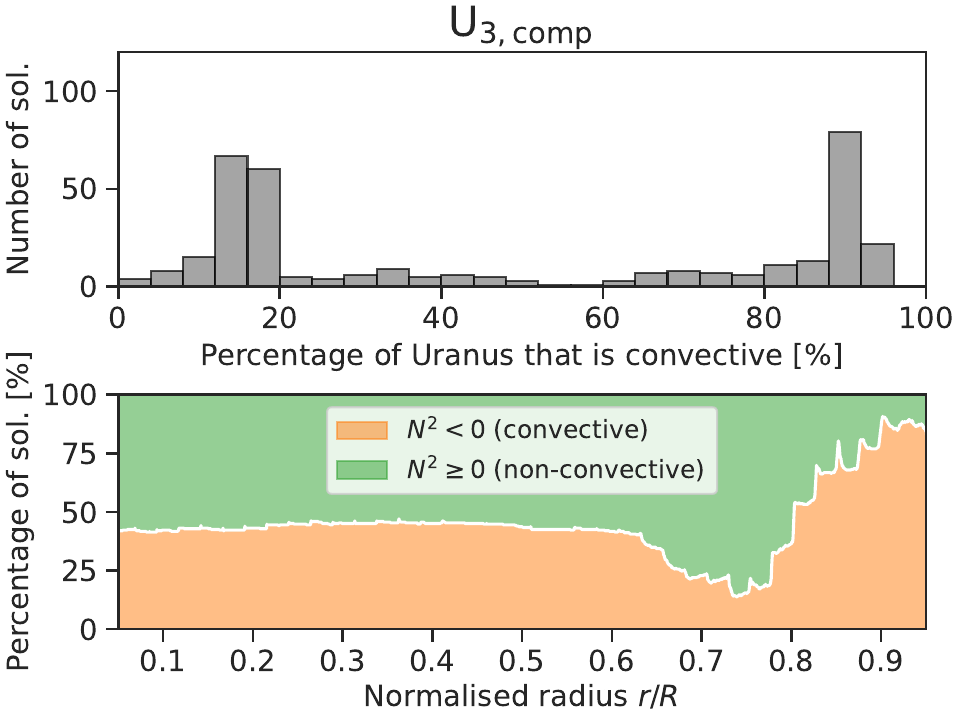}
    \includegraphics[width=\hsize]{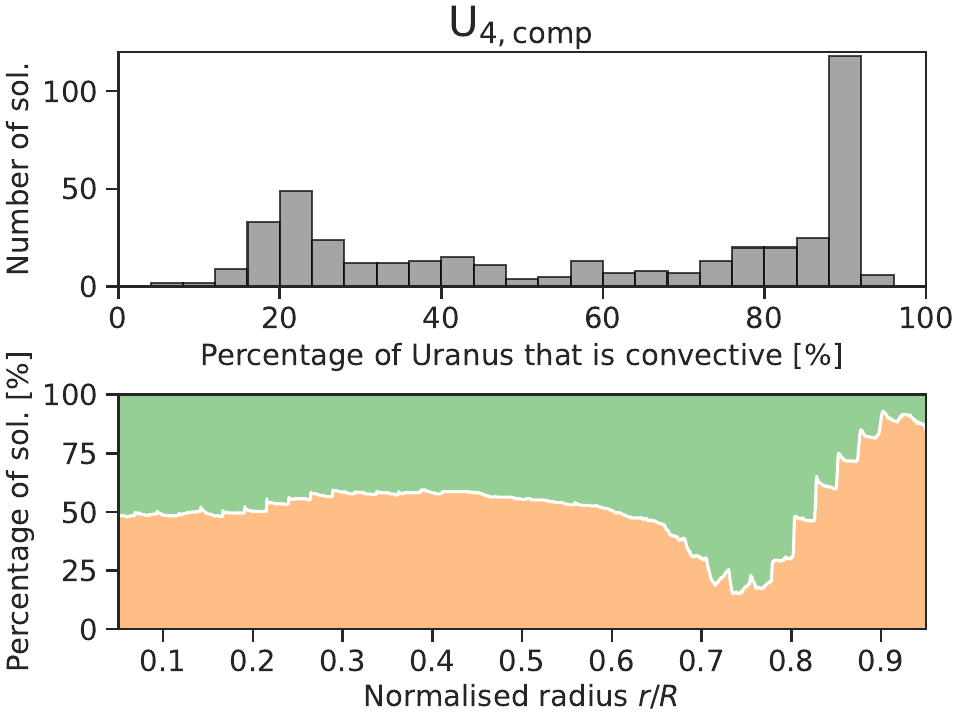}

    \caption{First and third plot: Histograms showing the percentage of Uranus' radius that is convective versus the number of solutions. In the U$_{4,\text{comp}}$ case, more than 100 empirical profiles were inferred to be convective for $\sim 90 \%$ of Uranus' radius. Second and last plot: Radial distributions of the convectivity for all solutions. In both cases, only $\sim 20$\% of the profiles were found to be convective at $\sim 0.75 r_{U}$.}
    \label{fig:Results_Convectivity}
    
\end{figure}

Figure \ref{fig:Results_Convectivity} summarises the convective regions inferred by our algorithm for the two cases presented in Figure \ref{fig:3UM4UM}. We begin by considering the histogram for the U$_{3,\text{comp}}$ case (top) and note that we can distinguish between two broad categories: Roughly half of our solutions are mostly non-convective ($\sim 20 \%$ of the planet w.r.t. its radius is convective) while the other half is mostly convective ($\sim 90 \%$ of the planet is convective). 
\par
In the U$_{4,\text{comp}}$ case (bottom), we can identify the same two categories of solutions in the histogram. However, there are more convective solutions overall. This is not surprising given the fact that there is more freedom to mix different elements, that is, more options to choose from for the U$_{4,\text{comp}}$ case. Some of these additional options are isentropic and compatible with convection, and since our algorithm is biased towards such solutions, we tend to find more convective solutions in the U$_{4,\text{comp}}$ case. The peaks of solutions that are convective for only $\sim 20 \%$ of the planet correspond to profiles possessing only a convective outermost part. The profile depicted in Figure \ref{fig:SampleResults_B} serves as an excellent example.
\par
Considering now the other plots, we see good agreement with the above discussion: Almost all profiles posses a convective outermost part. And almost all profiles possess a transition region around $\sim 0.75 r_{U}$. This seems to be a robust prediction given our 4-polytrope empirical models. The profiles then start to disagree below $\sim 0.75 r_{U}$, splitting up into the two aforementioned categories in roughly equally sized groups.

\subsection{Correlations}
\label{sec:Results_Correlations}

The random algorithm yielded many solutions for each empirical density profile, enabling the statistical analysis of the solutions. Here, we show the correlations of a few selected planetary properties for the U$_{4,\text{comp}}$ model: The mass fractions of different materials, central temperature and density, gravitational moments, and the mean location of the ionic H$_{2}$O regions that are convective. Unless specifically stated, the following correlations are also valid for the U$_{3,\text{comp}}$ model. The complete data for the U$_{3,\text{comp}}$ and U$_{4,\text{comp}}$ models are provided for a qualitative analysis in Appendix \ref{app:data_summary} (Figures \ref{fig:all_correlations2} and \ref{fig:all_correlations}). The stated p-values, which are all below 0.005, stay below 0.005 even when combining them to account for the Bonferroni correction\footnote{If multiple hypotheses are tested, the probability of observing a rare event increases and therefore, the likelihood of incorrectly rejecting a null hypothesis increases. The Bonferroni correction compensates for that increase by testing each individual hypothesis at a significance level adjusted to the number of hypotheses.}. 

\subsubsection*{Uranus' magnetic field generation}

\begin{figure}[H]
    \centering
    \includegraphics[width=\hsize]{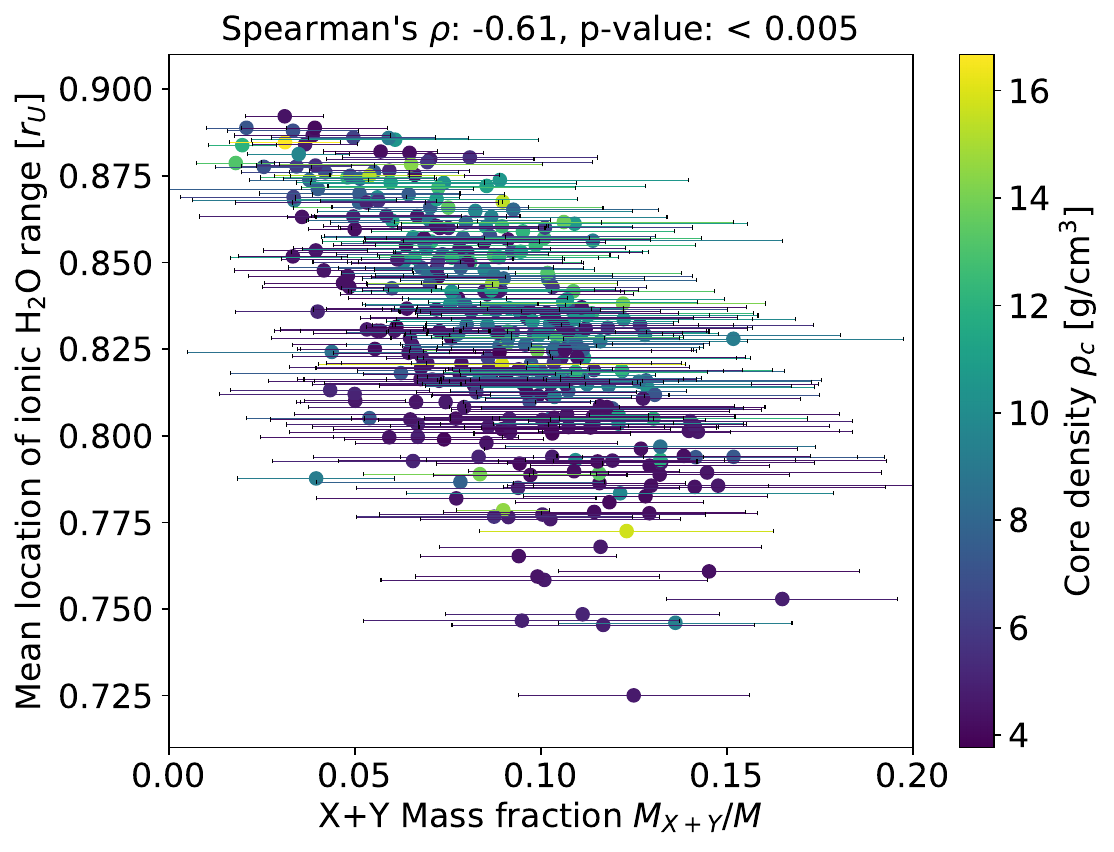}
    \caption{H-He mass in Uranus versus mean location of the convective ionic H$_{2}$O region. The error bars indicate the $1\sigma$ range of solutions found by the algorithm for density profiles possessing an ionic-water region. Profiles without a convective ionic H$_{2}$O region are not presented.}
    \label{fig:corr_ionic}
\end{figure}

Uranus has been observed to possess a magnetic field (\cite{Ness_1989}). Hence, interior modelling needs to explain this observation. According to Maxwell's equations, a magnetic field can be generated by moving charged particles. Given the right temperature and pressure conditions, water can become ionic (charged). Most of our models (77\% in the U$_{3,\text{comp}}$ case and 83\% in the U$_{4,\text{comp}}$ case) possess a convective region with ionic water. The convection would determine the velocity of the charged water particles generating the magnetic field. We now want to further investigate the properties of these regions.
First, we explore whether a higher abundance of a given element influences the location of the convective ionic water region. Figure \ref{fig:corr_ionic} shows the mean location of the convective ionic water region as a function of the total H-He mass fraction ($M_{X+Y}/M$). 

We find that a higher H-He mass fraction correlates with a deeper location of the convective ionic H$_{2}$O region. This can be explained by the larger extent of the H-He dominated outermost part of the planet, which pushes the H$_{2}$O rich regions further into Uranus' interior. Typically, the water mass fraction in our Uranus models is $\sim 0.2$ in the convective ionic H$_{2}$O regions. Some U$_{4,\text{comp}}$ models consist of a water mass fraction of $\sim 0.5$ and nearly $1$ for U$_{3,\text{comp}}$ models in these convective regions. The full data for the water mass fraction in convective ionic H$_{2}$O regions is presented in appendix \ref{app:data_summary}. At the moment, the minimum ionic water mass fraction that is required to ensure a high enough electrical conductivity to generate Uranus' magnetic field remains unknown \citep[for example][]{Soyuer_2020, Soderlund_2020}. Our work suggests that the inferred water mass fraction of $\sim 0.2$ is sufficient to explain Uranus' dynamo. Clearly, this topic should be investigated in future research in detail.
\par
It should be noted that not all of our U$_{4,\text{comp}}$ solutions possess a convective ionic H$_{2}$O region in the first place. We therefore only show the 83\% of solutions that possess one in Figure \ref{fig:corr_ionic}. For the extension of the convective ionic H$_{2}$O region, we found an upper limit of $\sim 0.15 r_{U}$ (see appendix \ref{app:data_summary}).

\subsubsection*{H-He in the deep interior of Uranus}

\begin{figure}[H]
    \centering
    \includegraphics[width=\hsize]{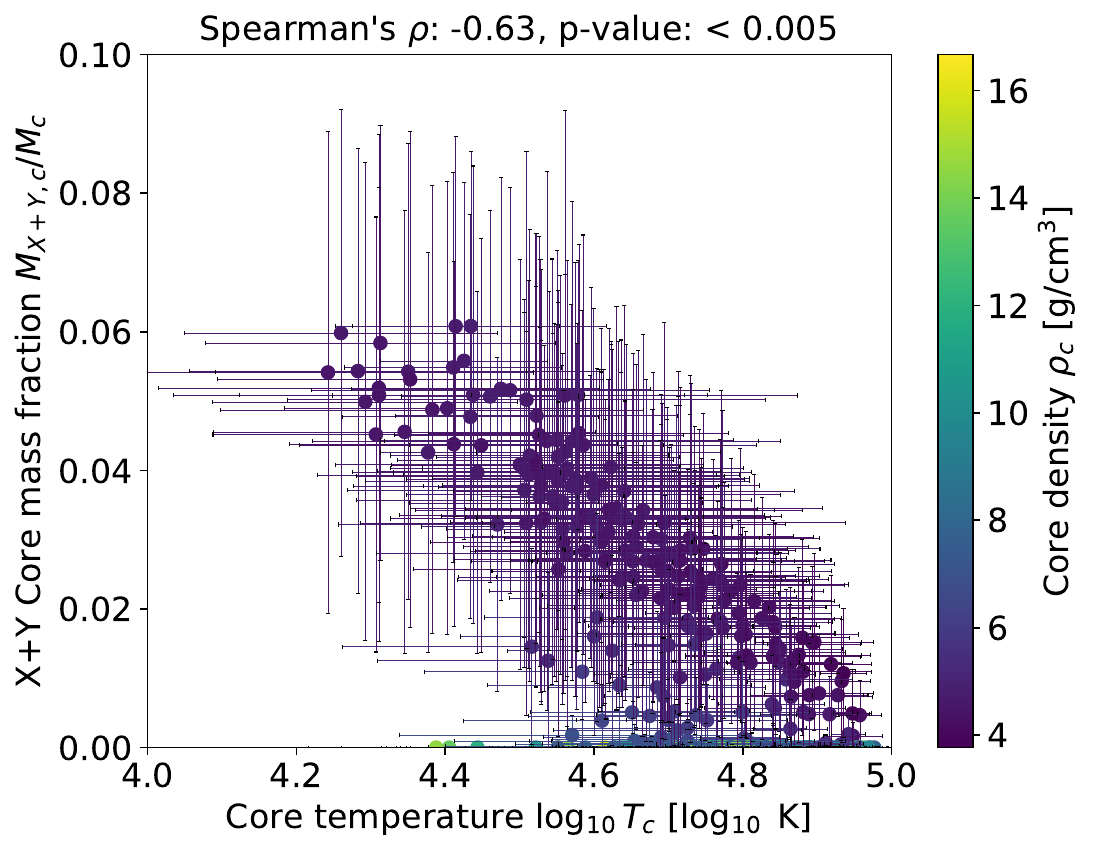}
    \caption{H-He mass fraction in Uranus' centre versus the central temperature $T_{c}$. The error bars indicate the $1\sigma$ range of solutions found by our algorithm.}
    \label{fig:corr_h_he_centre}
\end{figure}

Here we investigate the correlation between the centre H-He mass fraction $M_{X+Y,c}/M_{c}$ as a function of the central temperature $T_{c}$. The results are shown in Figure \ref{fig:corr_h_he_centre}. 

We find that the central temperature is inversely correlated with H-He abundance for lower central density empirical U$_{4,\text{comp}}$ models. Very high central temperatures are incompatible with a large H-He centre mass fraction. Overall, we find an upper limit of $\sim 0.1$ for the H-He mass fraction in Uranus' centre.  
Higher central density models do not include any H-He in the centre, since including H-He and Fe simultaneously as H$_{2}$O and SiO$_{2}$ is not allowed by construction (see Figure \ref{fig:3UM4UM}). U$_{3,\text{comp}}$ solutions could only find H-He in the central region if the central density is sufficiently low so it can be reproduced using a pure H-He-H$_{2}$O mixture.

\subsubsection*{The influence of future missions on Uranus' interior}
\label{sec:Results_AccurateJ6J8Measurements}

\begin{figure}[H]

    \centering
    \includegraphics[width=\hsize]{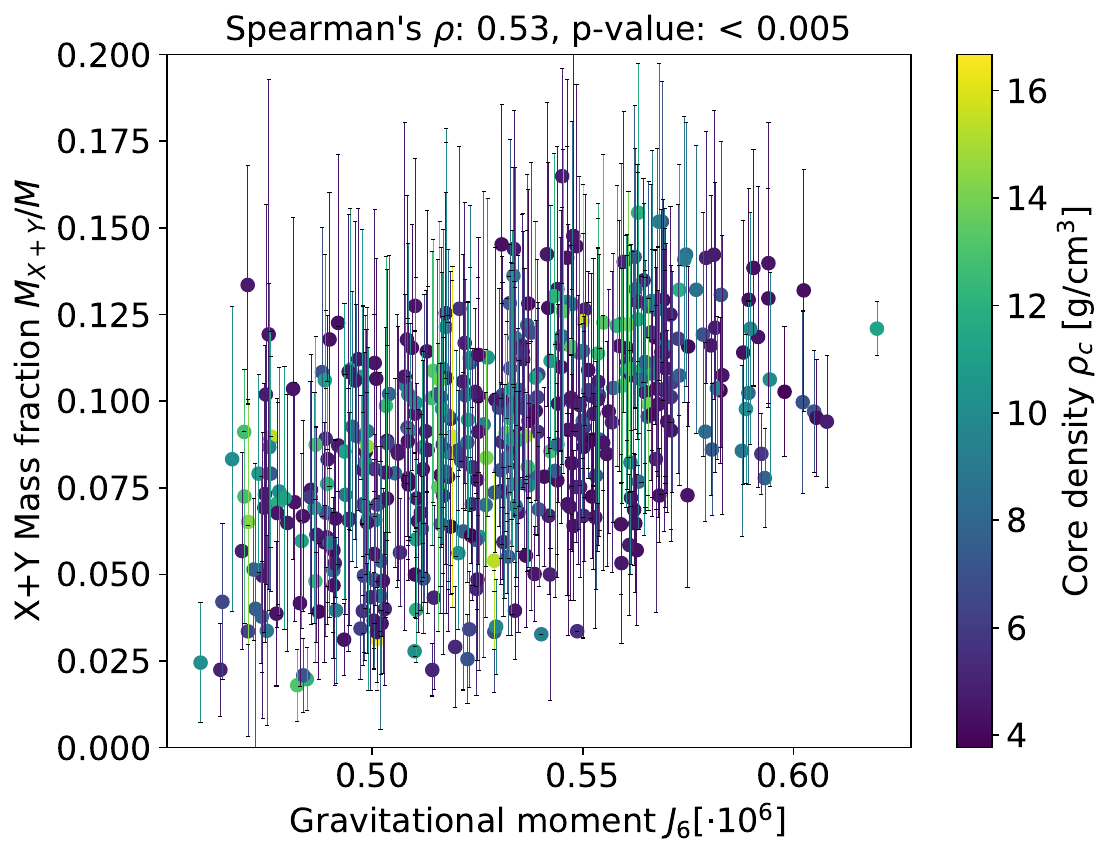}
    \includegraphics[width=\hsize]{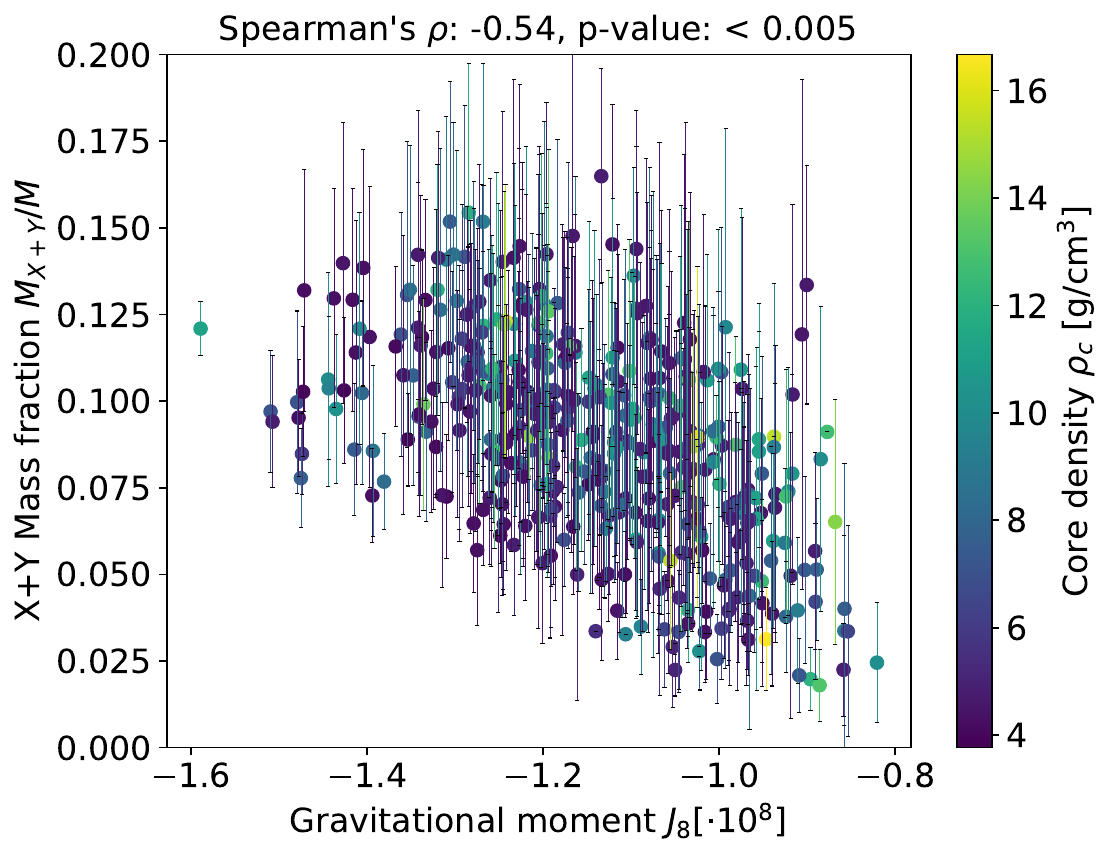}

    \caption{Total H-He mass fraction $M_{X+Y}/M$ versus the higher order gravitational moments $J_{6}$ (top) and $J_{8}$ (bottom). The errorbars indicate the $1\sigma$ range of solutions found by the random algorithm for each density profile.}
    \label{fig:corr_Js}
    
\end{figure}

Figure \ref{fig:corr_Js} shows the total H-He mass fraction $M_{X+Y}/M$ as a function of the higher order gravitational moments $J_{6}$ and $J_{8}$. We find clear correlations for both cases. We can therefore conclude that an accurate measurement of $J_{6}$ and $J_{8}$ can constrain $M_{X+Y}/M$ in Uranus: $J_{6}$ and $J_{8}$ are most sensitive in the outermost region of the planet, where H-He is most abundant. Since the mean location of the convective ionic H$_{2}$O region is also correlated with $M_{X+Y}/M$ (see Figure \ref{fig:corr_ionic}), accurate measurements of $J_{6}$ and $J_{8}$ can further constrain the location of the convective ionic water region. The fact that the spearman rank-order correlation coefficients are almost opposite to each other is no coincidence since the gravitational moments themselves are strongly correlated.

It is worth noting that a future mission will also more tightly constrain $J_2$ and $J_4$. Our work did not find significant correlations between the above quantities of interest and the gravitational moments $J_2$ and $J_4$.

\section{Discussion}
\label{sec:Discussion}

\subsection{Temperature sensitivity of the composition}
\label{sec:discussion_temperature}

In section \ref{sec:Results_Profiles}, we presented composition-temperature profiles and discussed that the homogeneous regions may exhibit a negative Brunt-Väisälä frequency too far away from zero. Another way to interpret this is that the inferred temperature profile is not adiabatic even though the region is homogeneous in composition and the algorithm labels it as a convective region. This occurs because empirical models can sometimes have a density gradient that is flatter than expected for an isentrope at a constant composition. This is true for empirical models in general and not just the ones presented here. This behaviour is noticeable when a physical EoS is used to interpret the composition and temperature of these models. Otherwise, it may remain hidden.
\par
In a region with a small density gradient, the two ways to match the low density of a deeper layer are to either increase the entropy (and leave the isentrope) or decrease the mean molecular weight of the mixture. Since the latter is not allowed in our algorithm, this can cause the entropy to increase inwards. As noted previously, this is not a stable configuration, because convection should quickly drive the temperature gradient towards an adiabatic one.
\par
\begin{figure}

    \centering
    \includegraphics[width=\hsize]{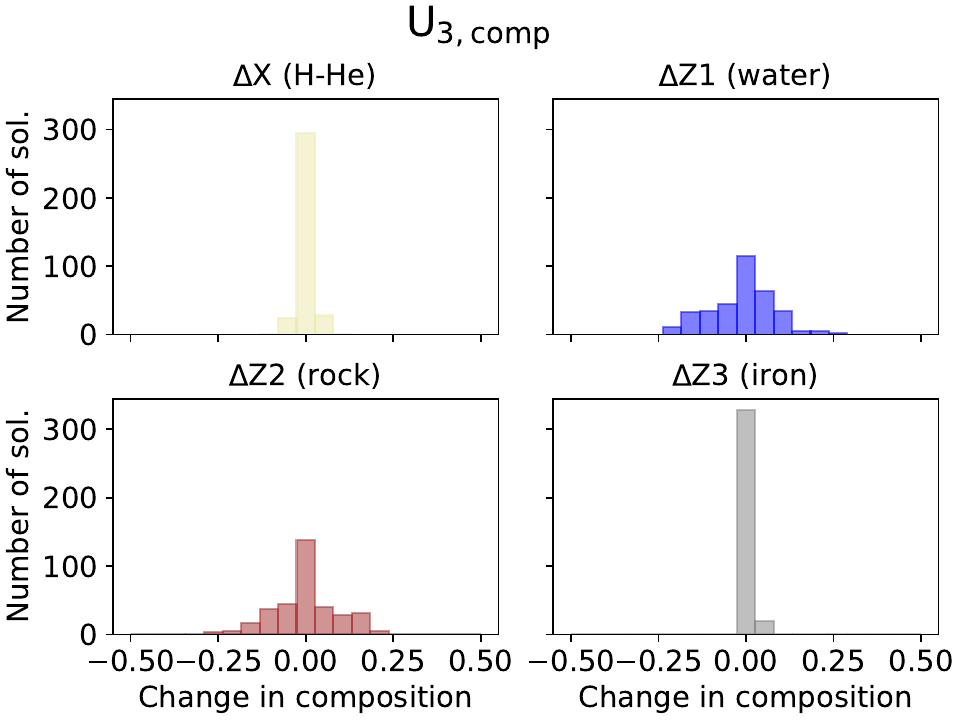}
    \includegraphics[width=\hsize]{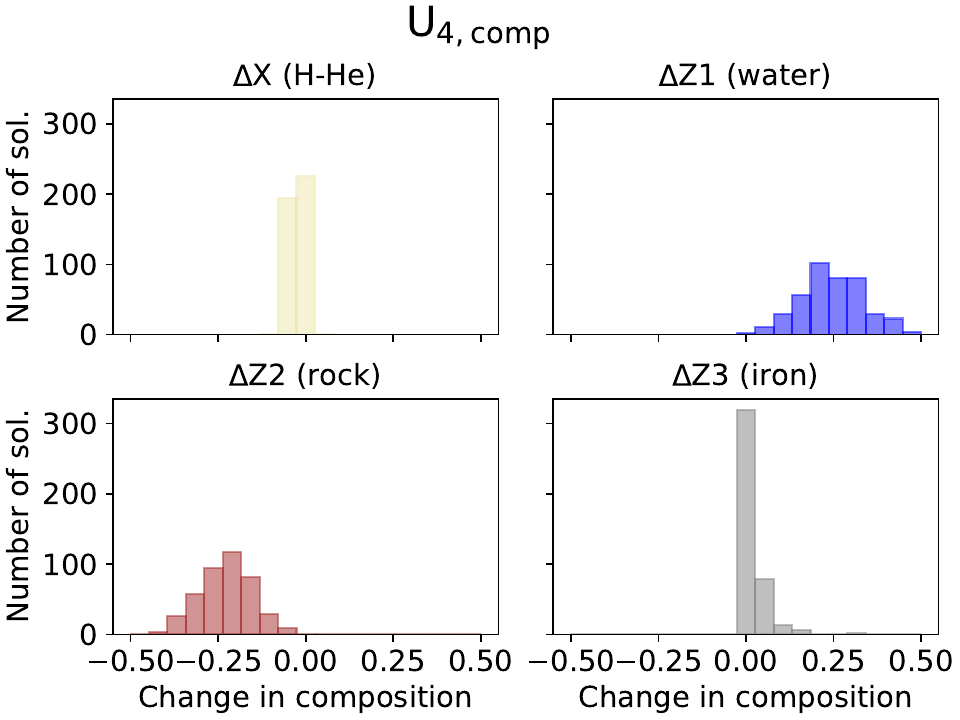}

    \caption{Absolute change in composition (H-He, water, rock, and iron) when enforcing a perfect adiabatic temperature profile. Positive numbers indicate that the forced adiabatic temperature profiles possess more of a given component. The results are shown for both U$_{3, \text{comp}}$ (top) and U$_{4, \text{comp}}$ (bottom) models.}
    \label{fig:delta_composition}
    
\end{figure}

In order to investigate how this issue affects our solutions, we forced an adiabatic temperature on the pressure-density profiles in regions labelled as convective by the algorithm. Forcing an adiabatic temperature gradient in convective regions leads to colder temperatures, the planet could have more lighter materials. While this approach is not fully self-consistent, it allows us to estimate the expected change in the inferred composition. We removed the constraint of an inwards monotonically increasing mean molecular weight to ensure that solutions are found despite the flat density gradients, for example. We then compared these new composition profiles with our previous results. Figure \ref{fig:delta_composition} shows the differences in composition for U$_{3, \text{comp}}$ (top) and U$_{4, \text{comp}}$ (bottom) models.

For the U$_{3, \text{comp}}$ models, the differences between the models are similar to Normal distributions peaked around zero. Therefore, despite the non-adiabatic temperature gradients in convective regions, we can conclude that our findings for the U$_{3, \text{comp}}$ are rather robust. On the other hand, for U$_{4, \text{comp}}$ models, the inferred composition differs more and follows a clear trend. While the changes in H-He and iron are small, the new inferred composition is biased towards more water and less rock. The magnitude of the shift is similar to the uncertainties shown in Figure \ref{fig:Results_Compositions_Easy}.
\par
\begin{figure}
    \centering
    \includegraphics[width=\hsize]{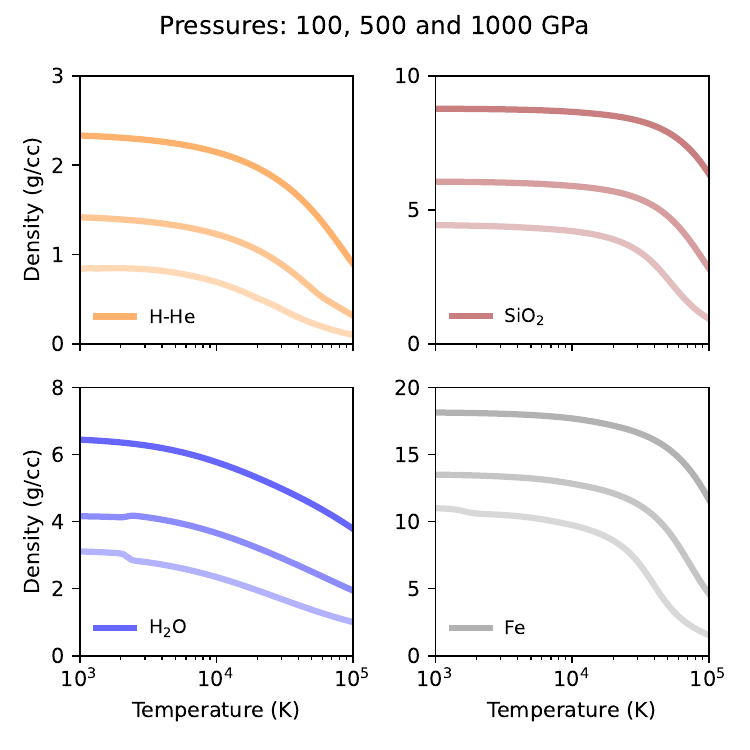}
    \caption{Density as a function of the temperature for three isobars at 100, 500, and 1000 GPa, with the pressure increasing from the bottom to the top of the figures. The four panels show the density-temperature relation of a protosolar H-He mixture, pure water, rock, and iron.}
    \label{fig:density_temperature_isobars_4panel}
\end{figure}

The changes in composition directly depend on the temperature. Figure \ref{fig:density_temperature_isobars_4panel} shows the temperature-density relation for a protosolar H-He mixture, pure water, rock, and iron. We used our EoS to calculate the density as a function of pressure at three different constant pressures of 100, 500, and 1,000 GPa. These pressures were chosen to represent the region inside the planet where most of the mass is.

As can be seen from the figure, at lower temperatures between about 1,000 to 10,000 K, the density does not strongly  depend  on the temperature, particularly for the more refractory rocks and iron. However, for temperatures beyond $\sim$10,000 K the density strongly decreases with temperature. As discussed in Section \ref{sec:Results_Profiles}, our models tend to be very hot, because of the composition gradients and the deviations from the isentropes in the convective layers. The combination of the hot interiors and the temperature-density relation causes the inferred composition to be sensitive to the temperature.
\par
Our test reveals that the inferred composition for the U$_{3, \text{comp}}$ should be robust, while this is less true for the U$_{4, \text{comp}}$ models, where we find a significant change of the water and rock mass fractions. To improve future constraints on the composition and temperature of Uranus, empirical models could restrict the polytropic indices so that density gradients do not become too flat, for example. We hope to address this in future studies.

\subsection{Comparison with previous work: Methods}

Our random algorithm differs in various ways compared to previous studies which focused on interpreting empirical models in terms of temperature and composition. For example, the algorithm of \cite{Neuenschwander_2024} starts in the outermost region of the planet, where the composition and temperature are reasonably well known. Then it moves towards the planetary centre layer by layer searching for isentropic, constant composition solution first. The algorithms are rather different when no isentropic and constant composition solutions are found: \cite{Neuenschwander_2024} numerically solve equation \ref{eq:N_squared} for $N^{2} = 0$ which leads to  a unique solution. This approach has two limitations: First, it can lead to very large temperature gradients if the composition changes rapidly. Second, it is unclear whether these solutions should be interpreted as marginally stable against large-scale convection or are essentially already unstable. On the other hand, our algorithm proposes numerous candidates that satisfies the inequality $N^{2} \geq 0$ and chooses randomly amongst them. The results of this study are therefore more agnostic in terms of the temperature gradient in non-convective layers. As a result,  we find a large range of possible temperature and composition profiles for a single empirical density profile. 
\par
The algorithm proposed by \cite{Podolak_2022} is rather different from the one we use here. As we discuss in detail in section \ref{sec:Introduction} it combines top-down and bottom-up approaches and minimises the amount of water layer by layer. Furthermore, their algorithm focuses on finding constant temperature rather than constant entropy solutions for both approaches. This leads to mostly isothermal solutions compared to our mostly isentropic solutions. As a result, they do not have information on entropy and can not identify the stability against convection in their models. 

\subsection{Comparison with previous work: Results}

Despite the above mentioned differences, our results for the internal structure and composition are consistent with these previous studies. \cite{Neuenschwander_2024} concluded that Uranus has a convective atmosphere and mantle on top of a non-convective inner region which is stable against convection. The transition between the convective and non-convective region was highly model-dependent and varied between $\sim40\%$ and $\sim85\%$ of Uranus’ radius. \cite{Neuenschwander_2024} found much higher central temperatures in comparison to standard adiabatic models. Our results can be viewed as an extension of those in \cite{Neuenschwander_2024}. We find numerous models that agree with the aforementioned description but also infer other type of solutions, such as almost purely convective interiors. This is due to the broader range of solutions found with our random algorithm. We conclude that the work presented in this paper is compatible with the results of \cite{Neuenschwander_2024} and shows even more possibilities for Uranus' internal structure. 
\par
\cite{Podolak_2022} found that the water-to-rock ratio in Uranus cannot be lower than 0.5 for their models that were analogously built to our U$_{3,\text{comp}}$ case. We confirm this trend as can be seen from Figure \ref{fig:Results_Compositions}: A low water-to-rock ratio is rare for U$_{3,\text{comp}}$ models and tends to require high central temperatures, as found by \cite{Podolak_2022}. We note that the models of \cite{Podolak_2022} were based on random density profiles which do not necessarily fit the available gravitational moments $J_{2}$ and $J_{4}$ (as only the moment of inertia was fitted). Despite the significantly different methods, the results are in agreement and should be considered to be robust.
\par
Finally, \cite{Malamud_2024} recently proposed an alternative to H-He being present in the core as a mechanism to explain a low water-to-rock ratio of Uranus and Neptune that is consistent with the expected composition of outer solar system planetesimals. They showed that chemical reactions between planetesimals dominated by organic-rich refractory materials and the hydrogen in gaseous atmospheres of protoplanets can form large amounts of methane. It was concluded that Uranus and Neptune could be compatible with having accreted refractory-dominated planetesimals, while still remaining 'icy' thanks to significant methane enrichment instead of water ice enrichment \citep[see][for details]{Malamud_2024} . 

\subsection{Suggested future investigations}

While our study presents an important advancement in empirical models, our work includes limitations and further improvements can be made in the future as listed below:

\begin{itemize}

    \item[$\bullet$] Our models are based on polytropic relations between the pressure and the density within the planet. A more agnostic approach would forego this for completely random density profiles that are consistent with the gravity data.
       
    \item[$\bullet$] Our models assume uniform rotation and no winds. In a refined study, the dynamical wind contributions to the gravity data should be accounted for. However, the difference due to winds on the overall composition is expected to be small  \citep{Neuenschwander_2024}.
    
    \item[$\bullet$] The atmospheric model \citep{Hueso_2020} and the EoS \citep{More_1988, Redmer_2011, Chabrier_2019} used for this work come with uncertainties themselves. These propagate directly to the results presented in this report.
    
    \item[$\bullet$] The $\beta$ pre-factor used for our algorithm was chosen ad hoc. It remains unclear what $\beta$ value would correspond to a physically realistic temperature gradient in non-convective regions. This problem directly propagates to the inferred central temperatures, which can be rather high.
    
    \item[$\bullet$] The order of magnitude of the Brunt-Väisälä frequency $N^{2}$ is unrealistically high. Future work should aim to decrease the typical order of magnitude further.
    
    
\end{itemize}

We suggest that a combination of both empirical (using randomness and relying primarily on gravity data) and physical (realistic EoS and density gradients) methods is a promising direction for future research. Such a semi-empirical approach would likely overcome several limitations of this work and provide both more robust and yet less biased predictions for planetary interior modelling.

\section{Summary and conclusions}
\label{sec:SummaryAndConclusion}

Empirical structure models have the advantage of probing solutions that are not covered by traditional interior models. First, we present new and improved empirical models of Uranus using four polytropes and tenth order ToF. These  models are improved compared to previous work due to the more physical representation of Uranus' atmosphere. The addition of the atmospheric polytrope increases the success rate of the random algorithm presented in this work by a factor of 10. 
The new random algorithm to infer the planetary temperature and composition profiles can account for two different structure models for Uranus. 
\par
We find that our empirical models would only match an adiabatic density gradient by an extraordinary coincidence. This is because the density profiles of empirical models are, by design, only made to fit the gravitational data. Consequently, in several cases we infer super-adiabatic temperatures that result in unrealistic Brunt-Väisälä frequencies. As we discussed earlier, in several cases differences in temperature can also affect the inferred planetary composition. We therefore suggest that the Brunt-Väisälä frequency could be used as an additional constraint when inferring the planetary composition from empirical models and we hope to address this topic in future research.
\par  
Overall, several key conclusions from our study can be made:
\begin{itemize}  

    \item[$\bullet$] The tenth order ToF provides accurate predictions of the gravitational moments up to $J_{14}$ assuming uniform rotation:
    $J_{6}  = ( 5.3078 \pm 0.3312)\cdot10^{-7}$,
    $J_{8}  = (-1.1114 \pm 0.1391)\cdot10^{-8}$,
    $J_{10} = ( 2.8616 \pm 0.5466)\cdot10^{-10}$,
    $J_{12} = (-8.4684 \pm 2.0889)\cdot10^{-12}$, and
    $J_{14} = ( 2.7508 \pm 0.7944)\cdot10^{-13}$.

    \item[$\bullet$]  Allowing H-He to be mixed into the deeper interior (as is expected from formation models) causes the inferred composition of Uranus to be rock-dominated: SiO$_{2}$ $\sim 56\%$, H$_2$O $\sim$ $33\%$, H-He $\sim 9$\%, and Fe $\sim2$\% of the total mass. 

    \item[$\bullet$]  We find an upper limit of $\sim 0.1$ for the H-He mass fraction in the centre.

    \item[$\bullet$]  No H-He in deeper parts of the planet leads to lower central temperatures and necessitates a higher H$_{2}$O abundance: H$_2$O $\sim$ $70\%$, SiO$_{2}$ $\sim 24\%$, H-He $\sim 6$\%, and Fe $\sim 1$\% of the total mass.

    \item[$\bullet$]  Almost all of our Uranus models are either mostly convective except for a few sharp transition regions for $\sim 90 \%$ of Uranus' radius $r_{U}$ or only convective in the outermost region above $\sim 0.8 r_{U}$. 

    \item[$\bullet$]  Almost all of our 4-polytrope based models possess a non-convective region around $\sim 0.75 r_{U}$. 

    \item[$\bullet$]  The majority of our models possess a region, ranging between $\sim (0.75-0.9) r_{U}$, that is both convective and possess ionic H$_{2}$O. This could explain the observed magnetic activity of Uranus. 

    \item[$\bullet$] The mean location of the convective ionic H$_{2}$O range is strongly correlated with the H-He mass fraction. It's length can range anywhere from barely above $0$, up to a maximum of $\sim 0.15 r_{U}$.  

    \item[$\bullet$] An accurate measurement of $J_{6}$ and $J_{8}$ would significantly constrain the composition solution space, most notably the H-He mass fractions and consequently the mean location of the convective ionic H$_{2}$O range. 

\end{itemize}

Our work suggests that a rock-dominated Uranus is a viable scenario when allowing the presence of H-He in its centre. We clearly show that Uranus' interior is complex and is likely to harbour composition gradients and non-adiabatic regions. This affects the inferred bulk composition and importantly the inferred water-to-rock ratio. These results should be linked to  formation and evolution models of Uranus. 
\par
Finally, we suggest that a dedicated space mission to Uranus is required to improve our understanding of Uranus' internal structure and bulk composition. Ideally, such a mission should measure Uranus' gravitational and magnetic field with high accuracy as well as the temperature and composition of its atmosphere. Such measurements will provide much-needed constraints to interior models that will eventually reveal key information on Uranus' interior, origin, and evolution.  

\begin{acknowledgements}
      The random algorithm presented in this work is an adapted version of the Master Thesis done by Sara Engeli at the Department of Astrophysics, University of Zurich. This work was supported by the Swiss National Science Foundation (SNSF) through a grant provided as a part of project number 215634: \url{https://data.snf.ch/grants/grant/215634}. We are thankful to the anonymous referee for valuable improvement suggestions and to Naor Movshovitz, his publicly available Theory of Figures code served as a foundation for our work.
\end{acknowledgements}

\bibliographystyle{aa}
\bibliography{literature.bib}

\begin{appendix}
\section{The Theory of Figures (ToF)}
\label{app:ToF}

Consider a planet with a total potential $U$. ToF (\cite{Zharkov_1978}) aims at characterising surfaces $r_l(\vartheta)$ of constant $U$, that is, finding solutions to the equation $U|_{r=r_l(\vartheta)} = \mathrm{constant}$. $r_l(\vartheta)$ parametrises the surface and is given by a spheroid Ansatz

\begin{equation}
r_l(\vartheta)=l\left(1+\sum_{n=0}^{\infty} s_{2 n}(l) P_{2 n}(\cos \vartheta)\right),
\end{equation}

where $\vartheta$ is the polar angle. $l$ denotes the average radius of the corresponding spheroid, that is

\begin{equation}
\label{eq:app_ToF3}
(4 \pi / 3) l^3=2 \pi \int_{-1}^1 d \cos \vartheta \int_0^{r_l(\vartheta)} d r^{\prime} r^{\prime 2}.
\end{equation}

Equation \ref{eq:app_ToF3} is called the condition of equal volume and can be used to calculate $s_{0}$ as a function of $s_{2}, s_{4}, \dots$. The Legendre polynomials are denoted by $P_{2 n}(\cos \vartheta)$. $s_{2 n}(l)$ are called figure functions and are to be determined. $U = V + Q$ is the sum of the Newtonian potential $V = -G \int d^{3}r^{\prime} \rho/|\boldsymbol{r^{\prime}}-\boldsymbol{r}|$ and the centrifugal potential $Q$, which is given by $Q = -\frac{1}{2}\omega^{2}r^{2}\sin^{2}(\vartheta)$ for the case of solid body rotation with a frequency $\omega$. By inserting $r_l(\vartheta)$ into $U(r) = V(r) + Q(r)$ and expanding every expression order by order one arrives at a general formula of the form

\begin{equation}
\label{eq:app_ToF_goal}
U(l, \vartheta)=-\frac{4 \pi}{3} G \bar{\rho} l^2 \sum_{k=0}^{\infty} A_{2 k}(l) P_{2 k}(\mu),
\end{equation}

where $\mu := \cos\vartheta$. The functions $A_{2 k}(l)$ are determined uniquely by the figure functions $s_{2n}(l)$, the frequency $\omega$, and the functions $S_{2n}(l)$ defined below. By assumption, it holds $U(l, \vartheta) = \mathrm{constant}$ as a function of $\vartheta$ on surfaces of constant total potential. Therefore, $A_{2 k}(l) = 0$ for $k>0$ and $U(l, \vartheta) = A_{0}(l)$ must hold on these surfaces. This conclusion will be crucial for the numerical algorithm described further down. The aforementioned expansion to arrive at equation \ref{eq:app_ToF_goal} is carried out by expanding the Newtonian potential

\begin{equation}
\label{eq:app_ToF1}
V(r, \vartheta)=-\frac{G}{r} \sum_{n=0}^{\infty}\left(r^{-2 n} D_{2 n}(r)+r^{2 n+1} D_{2 n}^{\prime}(r)\right) P_{2 n}(\mu),
\end{equation}

where we have introduced

\begin{align}
D_n(l) & =\frac{2 \pi}{n+3} \int_0^l d l^{\prime} \rho\left(l^{\prime}\right) \int_{-1}^1 d \mu^{\prime} P_n\left(\mu^{\prime}\right) \frac{d r^{n+3}}{d l}, \\
D_n^{\prime}(l) & =\frac{2 \pi}{2-n} \int_l^{R_{\mathrm{m}}} d l^{\prime} \rho\left(l^{\prime}\right) \int_{-1}^1 d \mu^{\prime} P_n\left(\mu^{\prime}\right) \frac{d r^{(2-n)}}{d l^{\prime}}, \quad(n \neq 2) \nonumber \\
D_2^{\prime}(l) & =2 \pi \int_l^{R_{\mathrm{m}}} d l^{\prime} \rho\left(l^{\prime}\right) \int_{-1}^1 d \mu^{\prime} P_2\left(\mu^{\prime}\right) \frac{d \ln r}{d l^{\prime}}. \nonumber
\end{align}

$R_{m}$ denotes the mean radius of the planet, that is $R_{m} = l_{\max}$. A dimensionless form can be obtained through the definitions

\begin{equation}
S_n(z)=\frac{3}{4 \pi \bar{\rho} l^{n+3}} D_n(l) \quad, \quad S_n^{\prime}(z)=\frac{3}{4 \pi \bar{\rho} l^{2-n}} D_n^{\prime}(l),
\end{equation}

At this point we have everything to get the functions $A_{2 k}(l)$ defined in equation \ref{eq:app_ToF_goal} as a function of the figure functions $s_{2n}(l)$, the frequency $\omega$, and the functions $S_{2n}(l)$. However, to implement and understand the numerical algorithm, we need to do additional work: By defining $r_{l}(\vartheta) = l(1+\Sigma(l,\vartheta))$ and

\begin{align}
S_n(z)&=\frac{1}{z^{n+3}} \int_0^z d z^{\prime} \frac{\rho\left(z^{\prime}\right)}{\bar{\rho}} \frac{d}{d z^{\prime}}\left[z^{\prime n+3} f_n\left(z^{\prime}\right)\right], \\
S_n^{\prime}(z)&=\frac{1}{z^{(2-n)}} \int_z^1 d z^{\prime} \frac{\rho\left(z^{\prime}\right)}{\bar{\rho}} \frac{d}{d z^{\prime}}\left[z^{\prime 2-n} f_n^{\prime}\left(z^{\prime}\right)\right], \nonumber \\
S_0(z)&=\frac{m(z)}{M z^3}, \nonumber
\end{align}

where 

\begin{align}
\label{eq:app_ToF2}
f_n(z) & =\frac{3}{2(n+3)} \int_{-1}^1 d \mu P_n(\mu)(1+\Sigma)^{n+3}, \\
f_n^{\prime}(z) & =\frac{3}{2(2-n)} \int_{-1}^1 d \mu P_n(\mu)(1+\Sigma)^{2-n} \quad(n \neq 2), \nonumber \\
f_2^{\prime}(z) & =\frac{3}{2} \int_{-1}^1 d \mu P_n(\mu) \ln (1+\Sigma), \nonumber
\end{align}

we can rewrite, after performing a partial integration and assuming $d\rho/dz$ to be finite,

\begin{align}
\label{eq:app_ToF4}
S_n(z)&=\frac{\rho(z)}{\bar{\rho}} f_n(z)-\frac{1}{z^{n+3}} \int_0^z \frac{d \rho}{\bar{\rho}} z^{\prime n+3} f_n\left(z^{\prime}\right), \\
S_n^{\prime}(z)&=-\frac{\rho(z)}{\bar{\rho}} f_n^{\prime}(z)+\frac{1}{z^{2-n}}\left(\frac{\rho(1)}{\bar{\rho}} f_n^{\prime}(1)-\int_z^1 \frac{d \rho}{\bar{\rho}} z^{\prime 2-n} f_n^{\prime}\left(z^{\prime}\right)\right). \nonumber
\end{align}

The integration over $\mu$ can be performed analytically for the expressions $f_n(z)$ and $f_n^{\prime}(z)$. We now posses convenient expressions for numerical evaluation: Divide the planet into a given number of spheroids $N$ with volumetric mean radii $l_1, \ldots, l_N$. Start with a (bad) guess for the value of the figure functions $s_{2n}$ at positions $l_i$ for $i \in \{1, \ldots,  N\}$. Equation \ref{eq:app_ToF2} then allows us to calculate the coefficients $f_{n}$ and $f^{\prime}_{n}$ which in turn gives us a guess for the functions $S_{n}$ and $S^{\prime}_{n}$ thanks to equation \ref{eq:app_ToF4}. Now one can check the conclusion from equation \ref{eq:app_ToF_goal}: If $A_{2k}(l_i)=0$ for $k>0$ and all $i \in \{1, \ldots,  N\}$ then the (bad) guess was actually good and the figure function values for surfaces of constant potential have been found. Otherwise, one can solve $A_{2k}(l)=0$ for $k>0$ explicitly to linear order \footnote{This is not as hard as might be expected, since the expressions $A_{2k}$ always have the form $A_{2k} = -s_{2k}S_{0} + [\ldots]$, therefore $s_{2k}=[\ldots]/S_{0}$ to linear order if $A_{2k}=0$.} to get a better guess for the figure functions $s_{2n}(l)$. This can be repeated until convergence is reached. The gravitational moments $J_{2n}$ can then be obtained by comparing equation \ref{eq:app_ToF1} with equation \ref{eq:ToF3} which yields 

\begin{equation}
    J_{2n} = - \left(\frac{R_m}{R_\text{eq}}\right)^{2n}S_{2n}(1).
\end{equation}

\subsection{Barotropic differential rotation}
\label{app:ToF_DiffRot}

We need to take a few steps back in order to understand what forms of differential rotation are allowed in the ToF discussion, that is, what kind of rotational potentials $Q$. We start with the assumptions used in \cite{Zharkov_1978} and consider a planet consisting of a fluid with a density $\rho$, pressure $p$ and velocity field $\vec{v}$. We use spherical coordinates $r$ (radius), $\vartheta$ (polar angle), and $\varphi$ (azimuthal angle). We assume that electromagnetic forces can be neglected and postulate

\begin{align}
    &\text{\small - Conservation of momentum for a static fluid with no viscosity:} \nonumber \\
    &(\vec{v} \cdot \vec{\nabla}) \vec{v} = -\frac{1}{\rho} \vec{\nabla} p + \vec{\nabla}V, \label{eq:NavStok} \\
    &\text{\small - Conservation of mass for a static fluid:} \nonumber \\
    &\vec{\nabla} \cdot ( \rho \vec{v} ) = 0, \label{eq:MassCons} \\
    &\text{\small - Newtonian gravity:} \nonumber \\
    &V(\vec{r}) = G \int \frac{\rho(\vec{r^{\prime}})}{|\vec{r}-\vec{r^{\prime}}|} \mathrm{d}\vec{r^\prime}, \label{eq:NewtGrav} \\
    &\text{\small - $p$ only depends on $\rho$, not the temperature $T$ (barotropic):} \nonumber \\
    &p = p(\rho), \label{eq:NoTemp} \\
    &\text{\small - $\vec{v}$ only has an azimuthal component $v_\varphi$ that is independent of $\varphi$:} \nonumber \\
    &\vec{v} = v_{\varphi}(r,\vartheta)\hat{e}_\varphi.\label{eq:VelocitySymm} 
\end{align}

Thanks to equation \ref{eq:NoTemp} we can define

\begin{equation}
    P(p) = \int_{p_0}^p \frac{\mathrm{d}p^{\prime}}{\rho(p^{\prime})} \implies \vec{\nabla}P = \frac{1}{\rho}\vec{\nabla}p
\end{equation}

due to the Fundamental Theorem of Calculus. Consequently, we are able to rewrite Equation \ref{eq:NavStok} as

\begin{equation}
    (\vec{v} \cdot \vec{\nabla}) \vec{v} = \vec{\nabla} \left( -P + V \right),
\end{equation}

which implies

\begin{equation}
    \label{eq:Curl_Zero}
    \vec{\nabla} \times \left((\vec{v} \cdot \vec{\nabla}) \vec{v}\right) = \vec{0}.
\end{equation}

Equation \ref{eq:Curl_Zero} is a testable condition for any velocity field $\vec{v}$ that is consistent with our assumptions. We can simplify equation \ref{eq:Curl_Zero} by using equation \ref{eq:VelocitySymm} for $(\vec{v} \cdot \vec{\nabla}) \vec{v}$:

\begin{align}
    \label{eq:VNablaV_res}
  (\vec{v} \cdot \vec{\nabla}) \vec{v} = \left(
    v_r \frac{\partial v_r}{\partial r}
  + \frac{v_\vartheta}{r} \frac{\partial v_r}{\partial \vartheta}
  + \frac{v_\varphi}{r\sin\vartheta} \frac{\partial v_r}{\partial \varphi}
  - \frac{v_\vartheta v_\vartheta + v_\varphi v_\varphi}{r}
  \right) &\hat{e}_r \\
+ \left(
    v_r \frac{\partial v_\vartheta}{\partial r}
  + \frac{v_\vartheta}{r} \frac{\partial v_\vartheta}{\partial \vartheta}
  + \frac{v_\varphi}{r\sin\vartheta} \frac{\partial v_\vartheta}{\partial \varphi}
  + \frac{v_\vartheta v_r}{r} - \frac{v_\varphi v_\varphi\cot\vartheta}{r}
  \right) &\hat{e}_\vartheta \nonumber \\
+ \left(
    v_r \frac{\partial v_\varphi}{\partial r}
  + \frac{v_\vartheta}{r} \frac{\partial v_\varphi}{\partial \vartheta}
  + \frac{v_\varphi}{r\sin\vartheta} \frac{\partial v_\varphi}{\partial \varphi}
  + \frac{v_\varphi v_r}{r}
  + \frac{v_\varphi v_\vartheta \cot\vartheta}{r}
  \right) &\hat{e}_\varphi \nonumber \\
  = \left( \frac{-v_\varphi^2}{r}\right) \hat{e}_r + \left( \frac{-v_\varphi^2\cot{\vartheta}}{r} \right) &\hat{e}_\vartheta. \nonumber
\end{align}

Since the curl of a vector field $\vec{A}$ in spherical coordinates is given by

\begin{align}
  \vec{\nabla} \times \vec{A} = \frac{1}{r\sin\vartheta} \left(
    \frac{\partial}{\partial \vartheta} \left(A_\varphi\sin\vartheta \right)
  - \frac{\partial A_\vartheta}{\partial \varphi}
  \right) &\hat{e}_r \\
{}+ \frac{1}{r} \left(
    \frac{1}{\sin\vartheta} \frac{\partial A_r}{\partial \varphi}
  - \frac{\partial}{\partial r} \left( r A_\varphi \right)
  \right) &\hat{e}_\vartheta  \nonumber \\
{}+ \frac{1}{r} \left(
    \frac{\partial}{\partial r} \left( r A_{\vartheta} \right)
  - \frac{\partial A_r}{\partial \vartheta}
  \right) &\hat{e}_\varphi, \nonumber
\end{align}

equation \ref{eq:Curl_Zero} states

\begin{align}
  \vec{\nabla} \times \left( (\vec{v} \cdot \vec{\nabla}) \vec{v} \right) = \frac{1}{r\sin\vartheta} \left(
    0
  - \frac{\partial}{\partial \varphi} \left( \frac{-v_\varphi^2\cot{\vartheta}}{r} \right)
  \right) &\hat{e}_r \\
{}+ \frac{1}{r} \left(
    \frac{1}{\sin\vartheta} \frac{\partial}{\partial \varphi} \left( \frac{-v_\varphi^2}{r}\right)
  - 0
  \right) &\hat{e}_\vartheta  \nonumber \\
{}+ \frac{1}{r} \left(
    \frac{\partial}{\partial r} \left( r \left( \frac{-v_\varphi^2\cot{\vartheta}}{r} \right) \right)
  - \frac{\partial}{\partial \vartheta} \left( \frac{-v_\varphi^2}{r}\right)
  \right) &\hat{e}_\varphi = \vec{0}. \nonumber
\end{align}

The $\hat{e}_r$ and $\hat{e}_\vartheta$ components are already zero thanks to the assumption that $v_{\varphi}(r,\vartheta)$ does not depend on $\varphi$. The $\hat{e}_\varphi$ component is not necessarily zero. Rather, it yields the condition

\begin{align}
    \label{eq:Cond}
    \frac{\partial}{\partial r} \left( r \left( \frac{-v_\varphi^2\cot{\vartheta}}{r} \right) \right)
    &= \frac{\partial}{\partial \vartheta} \left( \frac{-v_\varphi^2}{r}\right) \\ \Leftrightarrow \frac{\partial v_{\varphi}}{\partial r} \cot \vartheta &= \frac{\partial v_{\varphi}}{\partial \vartheta} \frac{1}{r} \nonumber
\end{align}

that quantifies velocity fields $\vec{v}$ that are acceptable given our assumptions. We now aim to define a total potential $U$ that facilitates the calculation of the pressure $p$ given a density $\rho$. Plugging in the result from equation \ref{eq:VNablaV_res} into equation \ref{eq:NavStok} component-wise, we arrive at

\begin{align}
    \frac{\partial p}{\partial r} &= \rho \left( \frac{v_\varphi^2}{r} + \frac{\partial V}{\partial r} \right) = \rho \frac{\partial }{\partial r} \left(\int \frac{v_\varphi^2}{r} dr +  V \right), \\
    \frac{1}{r} \frac{\partial p}{\partial \vartheta} &=  \rho \frac{1}{r}  \left( v_\varphi^2 \cot \vartheta + \frac{\partial V}{\partial \vartheta} \right) =  \rho \frac{1}{r} \frac{\partial }{\partial \vartheta} \left( \int v_\varphi^2 \cot \vartheta d \vartheta + V \right). \nonumber
\end{align}

Thanks to the condition derived in equation \ref{eq:Cond}, these two equations are actually one: $\vec{\nabla} p = \rho \vec{\nabla}(V+Q)$, where we defined

\begin{equation}
    \label{eq:QDefinition}
    Q = \int_{0}^{r} \frac{v_\varphi^2(r',\vartheta)}{r'} dr' = \int_{0}^{\vartheta} v_\varphi^2(r,\vartheta') \cot \vartheta' d\vartheta' .
\end{equation}

$U = V + Q$ is called the total potential. Surfaces of constant density, pressure and potential coincide, since $\vec{\nabla}p$ and $\vec{\nabla}U$ are pointing in the same direction (remember $\vec{\nabla} p = \rho \vec{\nabla}U$). Furthermore, $\vec{\nabla}p$ = $\rho \vec{\nabla}U \implies \vec{\nabla} \rho \times \vec{\nabla} U = 0$. So the normal vectors of $U$ and $\rho$ coincide at every point in space. One can solve

\begin{equation}
    \frac{\partial p}{\partial r} = \rho \frac{\partial U}{\partial r}
\end{equation}

to infer the pressure $p$ as a function of the radius in the planet as soon as the total potential $U$ and density $\rho$ are known. To summarise, we can indeed define $U = V + Q$ as a total potential and recover the condition for hydrostatic equilibrium, $\rho^{-1} \vec{\nabla} p = \vec{\nabla}U$, as long as equation \ref{eq:Cond} holds. Uniform rotation is characterised by $v_{\varphi} = \omega r \sin\vartheta$, satisfying equation \ref{eq:Cond} and yielding $Q = \frac{1}{2} \omega^2 r^2 \sin^2\vartheta$. A more general form is given by

\begin{equation}
    \label{eq:QDiffRot}
    Q = \omega^{2} r^{2} \sin^{2}\vartheta \left( \frac{1}{2} + \sum_{i=0}^{\infty} \frac{\alpha_{2i}}{2(i+1)}\frac{r^{2i}}{R_{m}^{2i}}\sin^{2i}\vartheta\right),
\end{equation}

and allows for higher order terms with free and unitless parameters $\alpha_{2i}$ deviating from uniform rotation. In the literature (for example \cite{Wisdom_2016}), $Q$ is often given as $Q = Q_0 + \Delta Q$, where $Q_0 = \omega^{2}r^{2}\sin^{2}(\vartheta)/2$ and

\begin{equation}
    \label{eq:Q_DR_Wisdom}
    \Delta Q = \sum_{i=1}^{n} C_{2i}\frac{r^{2i}}{R_{\text{eq}}^{2i}}\sin^{2i}\vartheta.
\end{equation}

The coefficients $C_{2i}$ have units of (length/time)$^2$. These two formulations are equivalent via the relation 

\begin{equation}
    \label{eq:Q_translation}
    \alpha_{2i} = \frac{2(i+1)R_{\text{m}}^{2i}C_{2(i+1)}}{\omega^2R_{\text{eq}}^{2(i+1)}}.
\end{equation}

\begin{figure}
    \centering
    \includegraphics[width=\hsize]{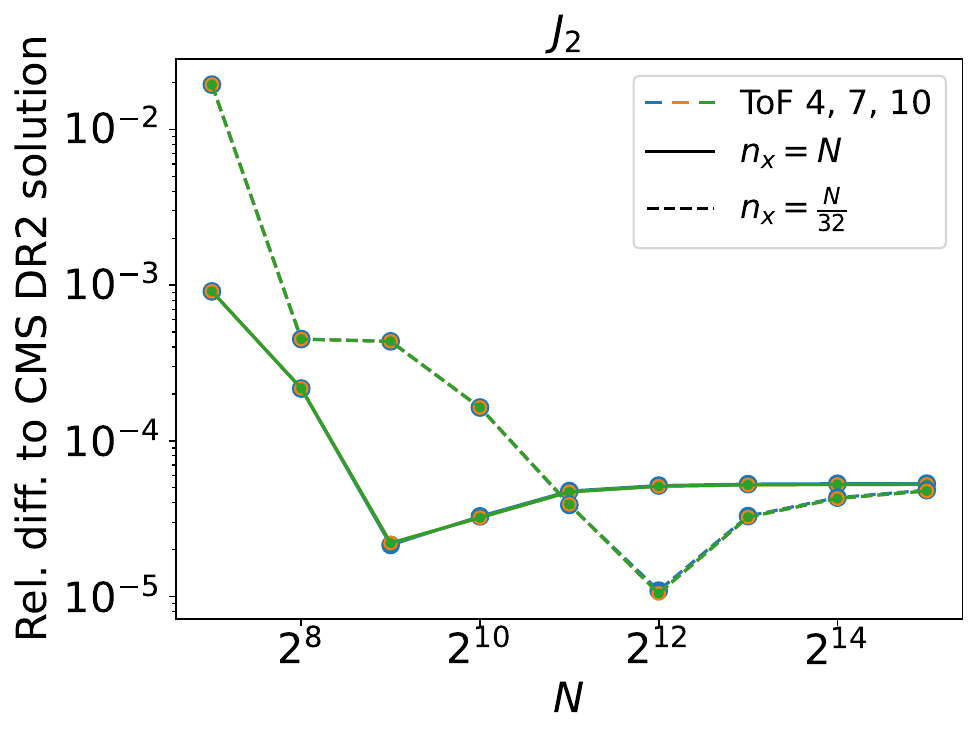}
    \includegraphics[width=\hsize]{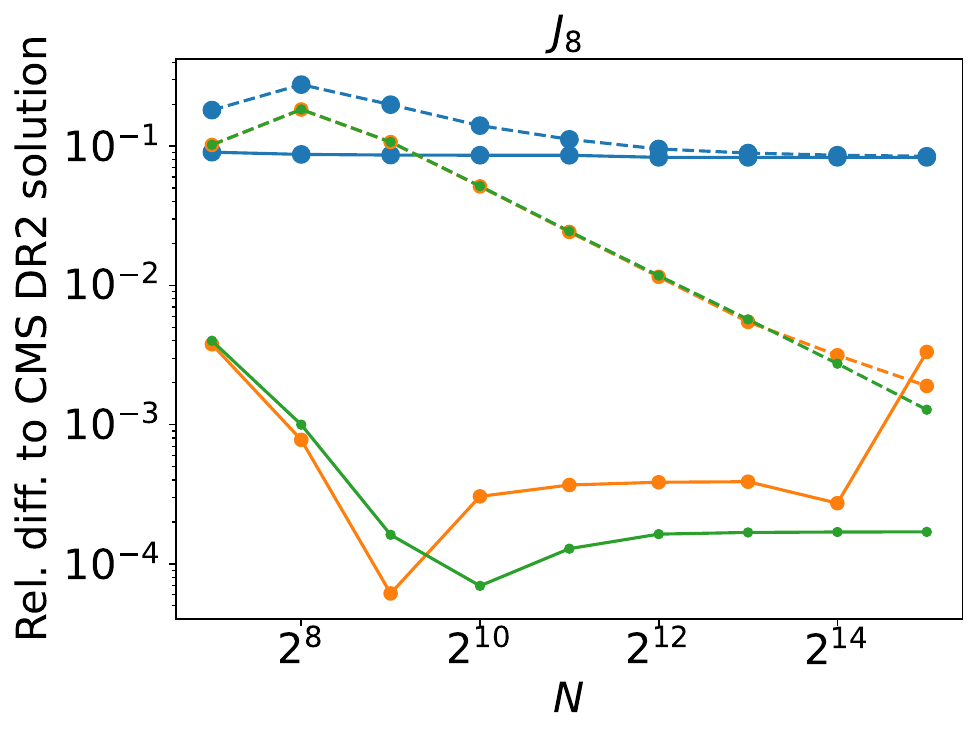}
    \caption{Relative differences to the CMS DR2 solutions by \cite{Wisdom_2016} of the gravitational moments $J_2$ (top) and $J_8$ (bottom) compared to our ToF solutions. We depict results for different numbers $N$ of used spheroids. We compare our implementation for different orders of the ToF and show the influence of using spline interpolation with the parameter $n_x$ that stands for the number of points that were calculated without interpolation.}
    \label{fig:CMSDR2_convergence}
\end{figure}

We now showcase the validity of our ToF implementation with barotropic differential rotation by comparing it to \cite{Wisdom_2016}. In addition to their simple Jupiter model with solid body rotation, \cite{Wisdom_2016} also presented results for a differentially rotating Jupiter. They called these models DR2 and DR3. We depict the relative difference for $J_{2}$ and $J_{8}$ between our implementation and \cite{Wisdom_2016} for the DR2 case in Figure \ref{fig:CMSDR2_convergence}. The DR3 case produces analogous results.

We find that, unlike in Figure \ref{fig:Bessel_convergence}, convergence is now stopped at a relative precision of 10$^{-4}$-10$^{-5}$. This is not indicative of a precision loss, but rather due to the fact that the CMS DR2 solutions are not perfectly accurate themselves. The CMS DR2 solutions were calculated using $2^9$ gridpoints and do not represent an analytical result. Our accuracy should hence exceed the accuracy of \cite{Wisdom_2016} for large enough $N$, which is consistent with a convergence halt that is visible in Figure \ref{fig:CMSDR2_convergence}. For future reference, we therefore state our DR2 and DR3 results (using tenth order ToF with $N=2^{15}$ spheroids and $n_x=N$) explicitly. The calculated gravitational moments beyond $J_{14}$ are increasingly inaccurate, a higher order ToF implementation would be required.

\begin{table}
\centering             
\caption{Calculated gravitational moments for the DR2 and DR3 barotropic differential rotation models of \cite{Wisdom_2016} using our tenth order ToF implementation with $N=2^{15}$ spheroids.} 
\begin{tabular}{l|l|l}
\hline
         & DR2                                          & DR3 \\
\hline
$J_{2}$  & $ \phantom{+} 1.399612111375 \cdot 10^{-2}$  & $ \phantom{+} 1.399672681976 \cdot 10^{-2}$ \\
$J_{4}$  & $ -           5.357561653756 \cdot 10^{-4}$  & $ -           5.358499109891 \cdot 10^{-4}$ \\
$J_{6}$  & $ \phantom{+} 3.158679614384 \cdot 10^{-5}$  & $ \phantom{+} 3.162687100692 \cdot 10^{-5}$ \\
$J_{8}$  & $ -           2.626134924416 \cdot 10^{-6}$  & $ -           2.714743623205 \cdot 10^{-6}$ \\
$J_{10}$ & $ \phantom{+} 2.744101342113 \cdot 10^{-7}$  & $ \phantom{+} 3.474673659768 \cdot 10^{-7}$ \\
$J_{12}$ & $ -           3.245006222520 \cdot 10^{-8}$  & $ \phantom{+} 1.880456488650 \cdot 10^{-8}$ \\
$J_{14}$ & $ \phantom{+} 4.096788490974 \cdot 10^{-9}$  & $ -           9.860482013894 \cdot 10^{-8}$ \\
$J_{16}$ & $ -           5.467528334579 \cdot 10^{-10}$ & $ \phantom{+} 1.075141393912 \cdot 10^{-7}$ \\
$J_{18}$ & $ \phantom{+} 6.960782064991 \cdot 10^{-11}$ & $ -           3.874804867708 \cdot 10^{-8}$ \\
$J_{20}$ & $ -           1.334484853106 \cdot 10^{-11}$ & $ -           3.156903298739 \cdot 10^{-8}$ \\
\hline
\end{tabular}%

\label{tab:DR2_DR3} 
\end{table}

Finally, we note that cubic spline interpolation is now significantly less accurate compared to the solid body rotation case depicted in Figure \ref{fig:Bessel_convergence}, even for high $N$.

\subsection{Mathematica file}

With this work, we provide a Mathematica file\footnote{\url{https://zenodo.org/doi/10.5281/zenodo.13340864}} that is able to calculate all ToF coefficients up to arbitrary order. However, the runtime increases exponentially with the order of the ToF. The Mathematica file is structured as follows:

\begin{itemize}
    \item The order up to which ToF should be calculated is defined with the parameter \texttt{ToFOrder}. 
    \item The order up to which differential rotation should be accounted for is defined with the parameter \texttt{DiffRotOrder}, that is
    \begin{align}
    Q = -\omega^{2} r^{2} \sin^{2}\vartheta & \Biggl( \frac{1}{2} \\
    &  + \sum_{i=1}^{\texttt{DiffRotOrder}} \frac{\alpha_{2i}}{2(i+1)}\frac{r^{2i}}{R_m^{2i}}\sin^{2i}\vartheta\Biggl). \nonumber
    \end{align}
    \item The expressions 
    \begin{equation}
        \label{eq:ToF1}
        (1+x)^{n+3}, (1+x)^{2-n}, \log(1+x), (\alpha+x)^{-n-1}, (\alpha+x)^{n}
    \end{equation}
    found in equations \ref{eq:app_ToF1} and \ref{eq:app_ToF2} are Taylor expanded up to order \texttt{ToFOrder}. $x$ and $\alpha$ are placeholders to be replaced later. 
    \item The first figure function $s_{0}$ is calculated as a function of $s_{2}, s_{4}, \dots$ based on the condition of equal volume (equation \ref{eq:app_ToF3}), we replace $s_{0} \rightarrow s_{0}(s_{2}, s_{4}, \dots)$.
    \item The surfaces of constant total potential
    \begin{align}
    \label{eq:ToF2}
        r_{l}(\vartheta) &= l \Biggl( 1 +s_{0}(s_{2}, s_{4}, \dots) \\ 
        &+ \sum_{n=1}^{\texttt{ToFOrder}} (\mathrm{OR})^{n} s_{2n}(l)P_{2n}(\cos\vartheta)\Biggr) 
        =: l \left( 1 + \Sigma_{l}(\mu)\right) \nonumber
    \end{align}
    are defined. OR is a helper parameter to keep track of the order of an expression. Subsequently, powers of $\Sigma_{l}(\mu)^{i}$ are calculated iteratively, only keeping terms of order (OR)$^{i}$ with $i \leq$ \texttt{ToFOrder}.
    \item The now calculated powers of $\Sigma_{l}(\mu)^{i}$ can be inserted into the Taylor expanded expressions from equation \ref{eq:ToF1}. More concretely, we replace $x^{i} \rightarrow (\Sigma_{l}(\mu)^{i}$ or $l^{i}\Sigma_{l}(\mu)^{i}$) and $\alpha^{i} \rightarrow l^{i}$ according to equations \ref{eq:app_ToF1} and \ref{eq:app_ToF2}. 
    \item Finally, the necessary integrations over the angle $\mu$ are performed to recover all coefficients for the numerical algorithm.   
\end{itemize}
    
\section{The Ledoux criterion}
\label{app:Ledoux}

In this section, we derive the form of the Ledoux criterion used in this work. In the following derivation, $g$ is the gravitational acceleration, and $\rho, P, T, S$ denote density, pressure, temperature, and entropy. We use the mean molecular weight $\mu$ to parameterise the composition. The gradients are defined as 

\begin{equation}
    \nabla_{ad} := \left( \frac{\partial \ln T}{\partial \ln P} \right)_{S}, \quad \nabla_T := \frac{d \ln T}{d \ln P}, \quad \nabla_{\mu} := \frac{d \ln \mu}{d \ln P} .
\end{equation}

We start with the Brunt-Väisälä frequency as in \cite{Kippenhahn_2013}

\begin{align}
    N^{2} &= g\frac{-\left( \frac{\partial \ln \rho}{\partial \ln T} \right)_{P,\mu}}{-\frac{dr}{d \ln P}}\left(\nabla_{ad} - \nabla_T \right) + g \frac{\left( \frac{\partial \ln \rho}{\partial \ln \mu} \right)_{P,T}}{-\frac{dr}{d \ln P}} \nabla_{\mu} \\
    &= \frac{g}{r}\frac{d \ln P}{d \ln r}\left( \frac{-1}{\left(\frac{\partial \ln T}{\partial \ln P}\right)_{\rho, \mu}\left( \frac{\partial \ln P}{\partial \ln \rho}\right)_{T, \mu}}\left(\nabla_{ad} - \nabla_T \right) - \left( \frac{\partial \ln \rho}{\partial \ln \mu} \right)_{P,T}\nabla_{\mu} \right)  \nonumber,
\end{align}

where we used the triple product rule

\begin{equation}
    -1 = \left( \frac{\partial \ln \rho}{\partial \ln T} \right)_{P, \mu} \left( \frac{\partial \ln T}{\partial \ln P} \right)_{\rho, \mu} \left( \frac{\partial \ln P}{\partial \ln \rho} \right)_{T, \mu}  .
\end{equation}

By defining 

\begin{equation}
    V := - \frac{d \ln P}{d \ln r}, \quad \chi_{\rho} := \left( \frac{\partial \ln P}{\partial \ln \rho} \right)_{T, \mu}, \quad \chi_{T} := \left( \frac{\partial \ln P}{\partial \ln T} \right)_{\rho, \mu}  ,
\end{equation}

we can write 

\begin{equation}
    N^{2} = \frac{gV}{r}\frac{\chi_{T}}{\chi_{\rho}}\left( \nabla_{ad} - \nabla_T + \frac{\chi_{\rho}}{\chi_{T}} \left( \frac{\partial \ln \rho}{\partial \ln \mu} \right)_{P,T} \nabla_{\mu} \right)  .
\end{equation}

In hydrostatic equilibrium $\rho^{-1}\vec{\nabla}P = \vec{\nabla}U$ and therefore $V$ is equal to $G M(r) \rho/(r P)$. With this we arrive at the following expression (see also \cite{Unno_1989}):

\begin{equation}
    N^{2} = \frac{g^{2}\rho}{P}\frac{\chi_{T}}{\chi_{\rho}}\left( \nabla_{ad} - \nabla_T + B \right)  ,
\end{equation}

where the composition term $B$ is

\begin{align}
    B &= \frac{\chi_{\rho}}{\chi_{T}}\left( \frac{\partial \ln \rho}{\partial \ln \mu} \right)_{P, T} \frac{d \ln \mu }{d \ln p} \\
    &=  \frac{\chi_{\rho}}{\chi_{T}}\left( \frac{\partial \ln \rho}{\partial \ln \mu} \right)_{P, T} \frac{d \ln \mu }{d \ln \rho} \frac{d \ln \rho }{d \ln p} \nonumber \\
    &= \frac{\chi_{\rho}}{\chi_{T}} \lim_{\delta \ln \rho \rightarrow 0} \frac{\ln \rho(P, T, \mu + \frac{d \mu}{d \ln \rho} \delta \ln \rho) - \ln\rho(P,T,\mu)}{\delta \ln \rho}\frac{d \ln \rho }{d \ln p} .\nonumber
\end{align}

The last line equality follows from

\begin{align}
    \ln \rho(P, T, \mu + \frac{d \mu}{d \ln \rho} \delta \ln \rho) =& \ln\rho(P,T, \mu) + \frac{d \mu}{d \ln \rho} \left( \frac{\partial \ln \rho }{\partial \mu} \right)_{P,T} \delta \ln \rho \nonumber \\
    &+ \mathcal{O}(\delta \ln \rho^{2}) \\
    =& \ln\rho(P,T, \mu) + \frac{d \ln \mu}{d \ln \rho} \left( \frac{\partial \ln \rho }{\partial \ln \mu} \right)_{P,T} \delta \ln \rho \nonumber \\
    &+ \mathcal{O}(\delta \ln \rho^{2}) \nonumber .
\end{align}

Therefore, $B$ can be evaluated numerically as

\begin{align}
    B &= \frac{\chi_{\rho}}{\chi_{T}} \lim_{\delta \ln \rho \rightarrow 0} \frac{\ln \rho(P, T, \mu + \frac{d \mu}{d \ln \rho} \delta \ln \rho) - \ln\rho(P,T,\mu)}{\delta \ln \rho}\frac{d \ln \rho }{d \ln p} \nonumber \\
    &= \frac{\chi_{\rho}}{\chi_{T}}\frac{\ln \rho \left(P_{n}, T_{n}, \mu_{n+\Delta n} \right) - \ln \rho \left(P_{n}, T_{n}, \mu_{n} \right)}{\ln \rho_{n+\Delta n} - \ln \rho_{n}} \frac{\ln \rho_{n+\Delta n} - \ln \rho_{n}}{\ln P_{n+\Delta n} - \ln P_{n}} \nonumber \\
    &= \frac{\chi_{\rho}}{\chi_{T}}\frac{\ln \rho \left(P_{n}, T_{n}, \mu_{n+\Delta n} \right) - \ln \rho \left(P_{n}, T_{n}, \mu_{n} \right)}{\ln P_{n+\Delta n} - \ln P_{n}} .
\end{align}

\section{Further data and results}
\label{app:data_summary}

\begin{figure}
    \centering
    \includegraphics[width=\hsize]{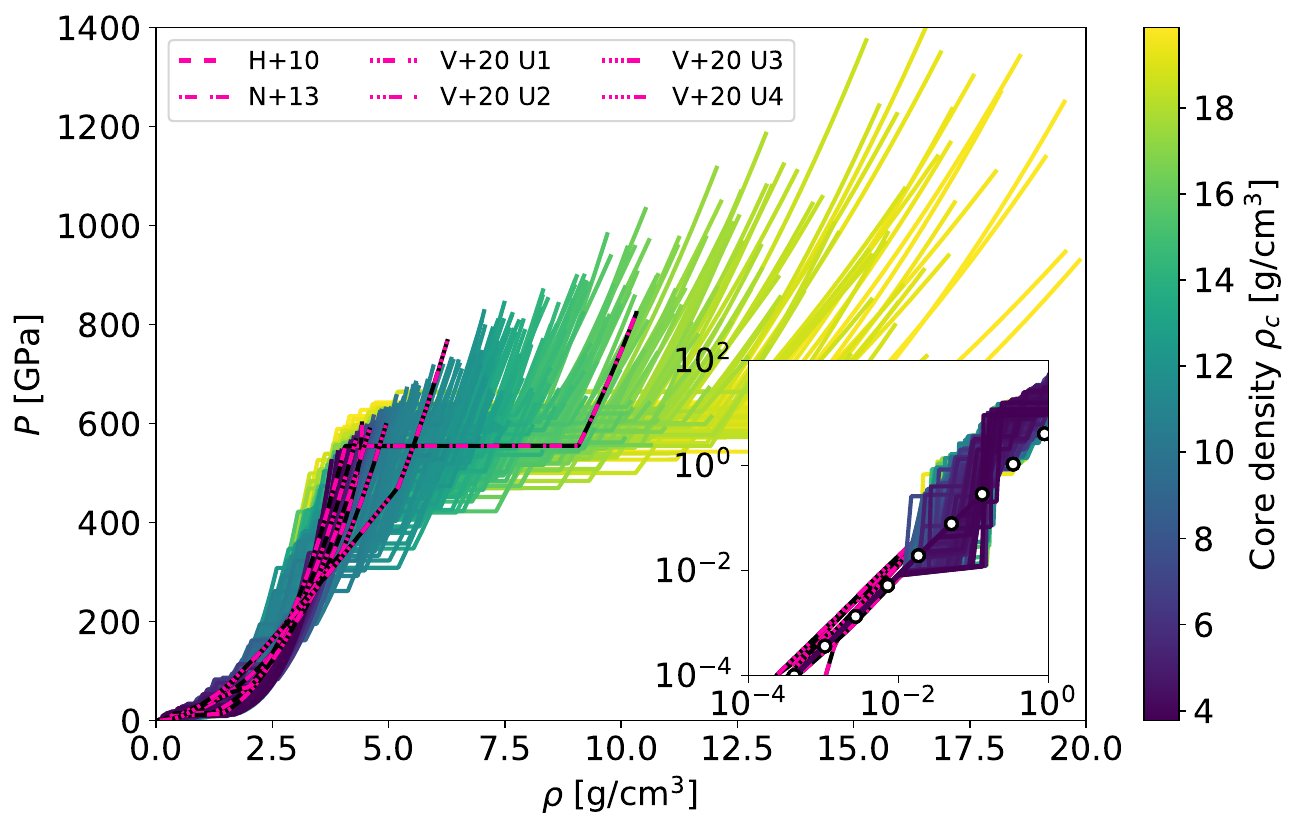}
    \caption{Uranus' pressure $P$ as a function of the density $\rho$. Other pressure-density profiles that have been published by previous studies are also presented for comparison \citep{Helled_2010,Nettelmann_2013,Vazan_2020}. The panel on the right shows the outermost region. The white dots show the atmosphere model by \cite{Hueso_2020}.}
    \label{fig:P_of_rho}
\end{figure}

Figure \ref{fig:P_of_rho} shows the pressure $P$ of our empirical interior models as a function of their density $\rho$. We depict the same zoomed-in subplot that is used in Figure \ref{fig:rho_of_r} and now include the whole solution range.

\begin{figure}
    \centering
    \includegraphics[width=\hsize]{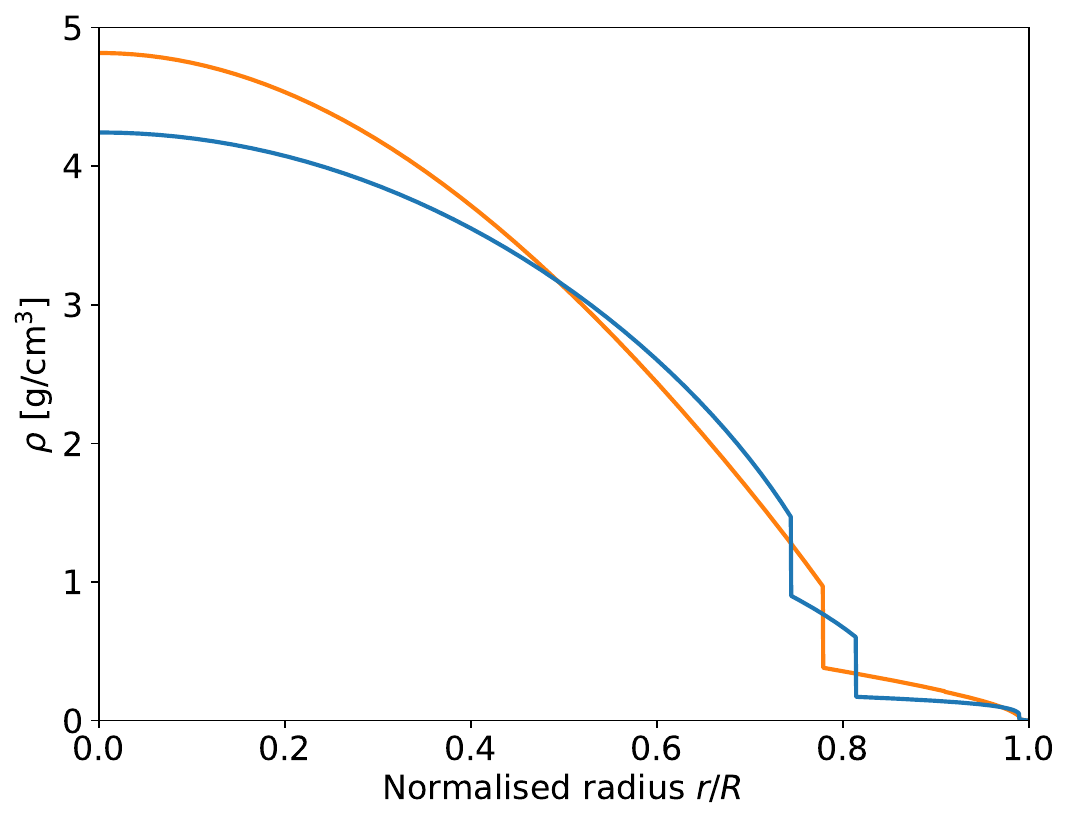}
    \caption{Direct comparison of the density profiles depicted in Figures \ref{fig:SampleResults_A} (blue) and \ref{fig:SampleResults_B} (orange).}
    \label{fig:compare30vs160}
\end{figure}

Figure \ref{fig:compare30vs160} displays a direct comparison of the density profiles depicted in Figures \ref{fig:SampleResults_A} (blue) and \ref{fig:SampleResults_B} (orange).

We find that although the density profiles are indeed somewhat similar, they are distinct enough to justify the fact that the random algorithm infers significantly different temperature and composition profiles for the two of them. The blue density profile (from Figure \ref{fig:SampleResults_A}) has a core density that is $\sim15\%$ lower and a less steep density gradient compared to the orange density profile (from Figure \ref{fig:SampleResults_B}).

\begin{figure*}
\centering
\resizebox{\hsize}{!}{
\includegraphics{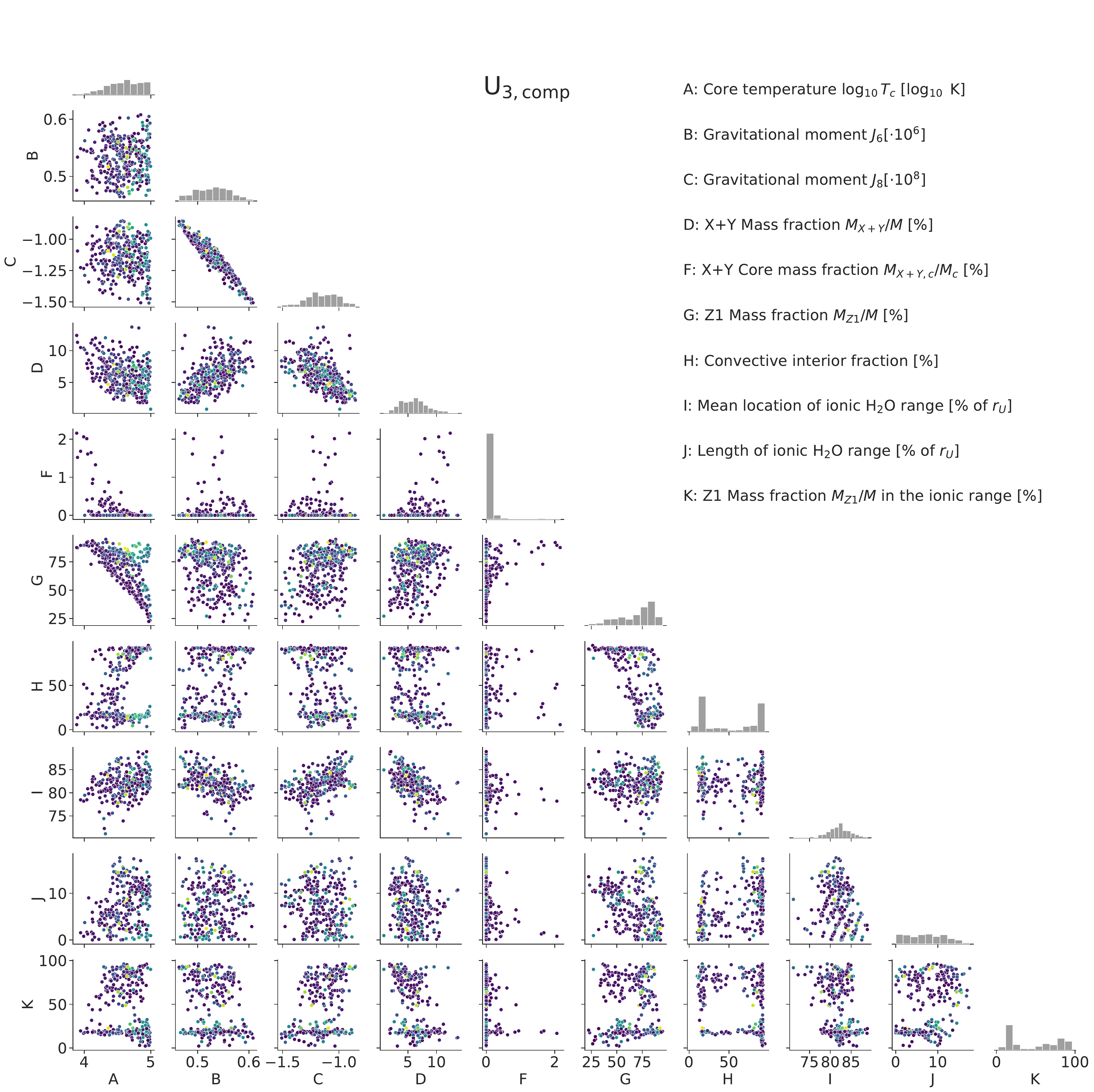}
}
\caption{Summarised results of the random algorithm for 4-polytrope empirical models of Uranus (U$_{3,\text{comp}}$ case). Each individual point represents one empirical profile. There are no errorbars indicating the range for a single profile to enhance readability.}
\label{fig:all_correlations2}
\end{figure*}

\begin{figure*}
\centering
\resizebox{\hsize}{!}{
\includegraphics{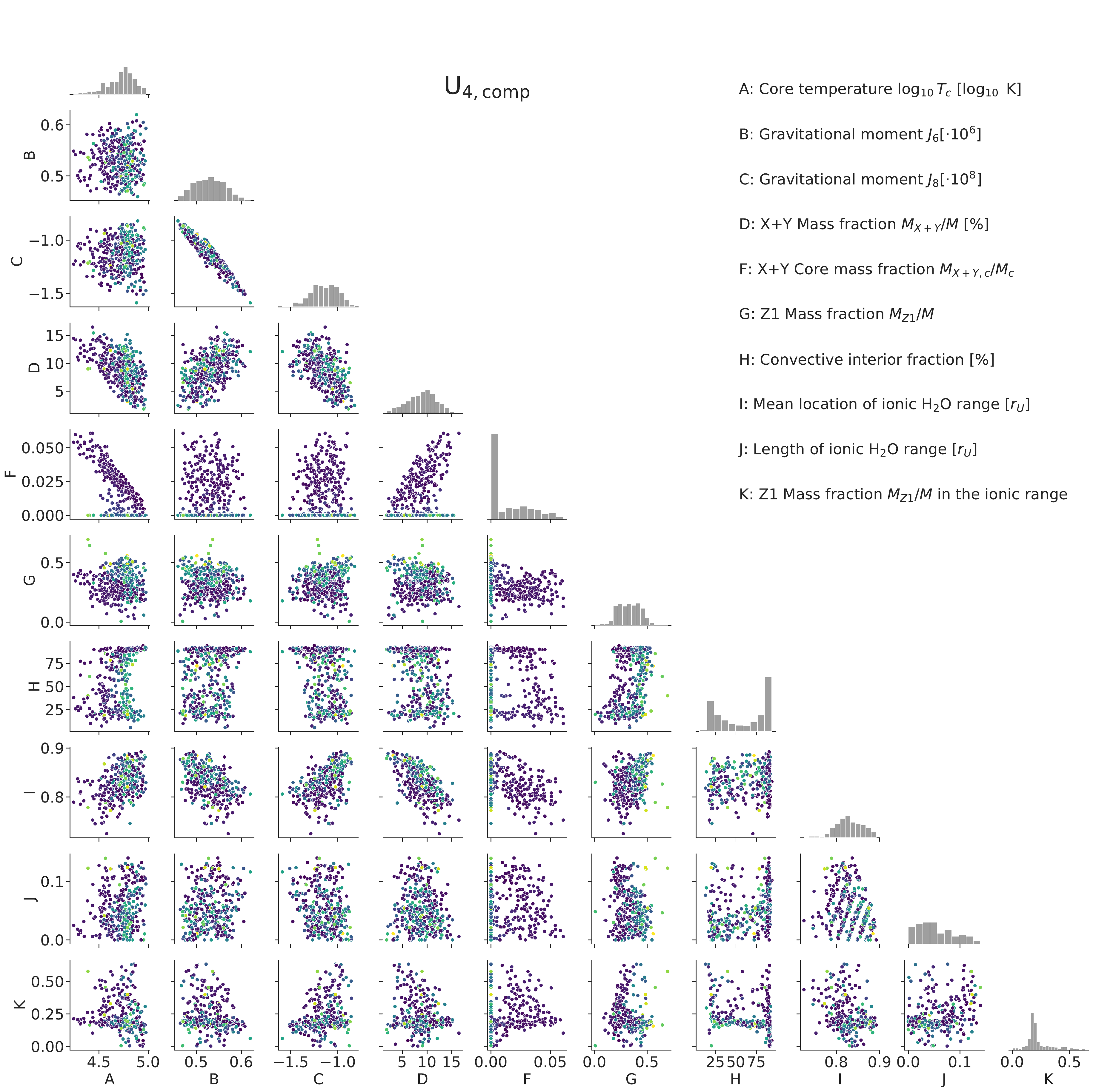}
}
\caption{Summarised results of the random algorithm for 4-polytrope empirical models of Uranus (U$_{4,\text{comp}}$ case). Each individual point represents one empirical profile. There are no errorbars indicating the range for a single profile to enhance readability.}
\label{fig:all_correlations}
\end{figure*}

Figures \ref{fig:all_correlations2} and \ref{fig:all_correlations} present all correlation plots of our results for the U$_{3,\text{comp}}$ and U$_{4,\text{comp}}$ case, respectively. The plots involving the convective ionic H$_{2}$O range only include models that posses a convective ionic H$_{2}$O region in the first place. These are 77\% and 83\% of solutions, respectively.

\end{appendix}

\end{document}